\documentclass[sigconf]{acmart}

\usepackage{xspace}
\usepackage{enumitem}
\usepackage{pifont}
\usepackage{makecell}
\usepackage{diagbox}
\usepackage{graphicx}
\usepackage{verbatim}
\usepackage{multirow}

\usepackage[ruled, lined, linesnumbered, commentsnumbered, longend]{algorithm2e}

\SetCommentSty{mycommfont}

\usepackage{hyperref}
\usepackage{longtable}
\usepackage{array}
\usepackage{subfigure}
\usepackage{stfloats}
\usepackage{multicol}
\usepackage{color}
\usepackage{epstopdf}
\usepackage{bm}
\usepackage{amsmath}
\usepackage{booktabs} 
\usepackage{epsfig}
\usepackage{enumitem}
\usepackage{cleveref}
\usepackage{arydshln}
\usepackage{balance}
\usepackage{xspace}
\usepackage{enumitem}
\usepackage{listings}
\usepackage{xcolor}

\usepackage{caption}
\usepackage{subcaption}

\newcommand{\hide}[1]{} 
\newcommand{\beq}[1]{\vspace{-0.03in}\begin{equation}#1\end{equation}\vspace{-0.03in}}
\newcommand{\vpara}[1]{\vspace{0.05in}\noindent \textbf{#1 }}

\newcommand{\model}{BOND\xspace}
\newcommand{\da}{WhoIsWho\xspace }

\newtheorem{problem}{Problem}

\AtBeginDocument{%
  \providecommand\BibTeX{{%
    \normalfont B\kern-0.5em{\scshape i\kern-0.25em b}\kern-0.8em\TeX}}}

\copyrightyear{2024}
\acmYear{2024}
\setcopyright{acmlicensed}
\acmConference[WWW '24]{Proceedings of the ACM Web Conference 2024}{May 13--17, 2024}{Singapore, Singapore}
\acmBooktitle{Proceedings of the ACM Web Conference 2024 (WWW '24), May 13--17, 2024, Singapore, Singapore}
\acmDOI{10.1145/3589334.3645580}
\acmISBN{979-8-4007-0171-9/24/05}

\acmSubmissionID{1585}

\begin{document}

\title{\model: Bootstrapping From-Scratch Name Disambiguation with Multi-task Promoting}


\author{Yuqing Cheng}
\authornote{Equal contribution.}
\authornote{Work was done when Yuqing interned at Zhipu AI.}
\orcid{0009-0003-3634-1202}
\affiliation{%
  \institution{Central Conservatory of Music}
  \city{Beijing}
  \country{China}
  }
\email{chengyuqing@mail.ccom.edu.cn}

\author{Bo Chen}
\authornotemark[1]
\orcid{0000-0002-9629-5493}
\affiliation{%
  \institution{Tsinghua University}
  \city{Beijing}
  \country{China}
  }
\email{cb21@mails.tsinghua.edu.cn}

\author{Fanjin Zhang}
\authornotemark[1]
\authornote{Fanjin Zhang and Jie Tang are the corresponding authors.}

\orcid{0000-0001-8551-1966}
\affiliation{%
  \institution{Tsinghua University}
  \city{Beijing}
  \country{China}
  }
\email{fanjinz@tsinghua.edu.cn}

\author{Jie Tang}
\authornotemark[3]
\orcid{0000-0003-3487-4593}
\affiliation{%
  \institution{Tsinghua University}
  \city{Beijing}
  \country{China}
  }
\email{jietang@tsinghua.edu.cn}

\renewcommand{\shortauthors}{Yuqing Cheng, Bo Chen, Fanjin Zhang, and Jie Tang}

\begin{abstract}
From-scratch name disambiguation is an essential task for establishing a reliable foundation for academic platforms. It involves partitioning documents authored by identically named individuals into groups representing distinct real-life experts.
Canonically, the process is divided into two decoupled tasks: locally estimating the pairwise similarities between documents followed by globally grouping these documents into appropriate clusters. 
However, such a decoupled approach often inhibits optimal information exchange between these intertwined tasks.
Therefore, we present \model, which bootstraps the local and global informative signals to promote each other in an end-to-end regime.
Specifically, \model harnesses local pairwise similarities to drive global clustering, subsequently generating pseudo-clustering labels. These global signals further refine local pairwise characterizations.
The experimental results establish \model's superiority, outperforming other advanced baselines by a substantial margin.
Moreover, an enhanced version, BOND+, incorporating ensemble and post-match techniques, rivals the top methods in the WhoIsWho competition\footnote{\url{http://whoiswho.biendata.xyz/}}. 
\end{abstract}



\hide{
\received{20 February 2007}
\received[revised]{12 March 2009}
\received[accepted]{5 June 2009}
}

\maketitle

\section{Introduction}

\label{sec:intro}

Name disambiguation is a core component in online academic systems such as Google Scholar, DBLP, and AMiner~\cite{tang2008arnetminer}. With the exponential growth of research documents in recent years~\cite{zhang2019oag}, the problem of author name ambiguity has become more complex. This issue encompasses scenarios where identical authors exhibit diverse name variations, distinct authors share identical names, or instances of homonyms.
For instance, as of October 2023, DBLP contained over 300 author profiles with the name "Wei Wang" in the field of computer science alone, not to mention across all academic disciplines. This underscores the pressing demand for the development of efficient and scalable algorithms tailored to confront the challenges presented by author name ambiguity.

\begin{figure}
    \centering
    \includegraphics[width=0.45\textwidth]{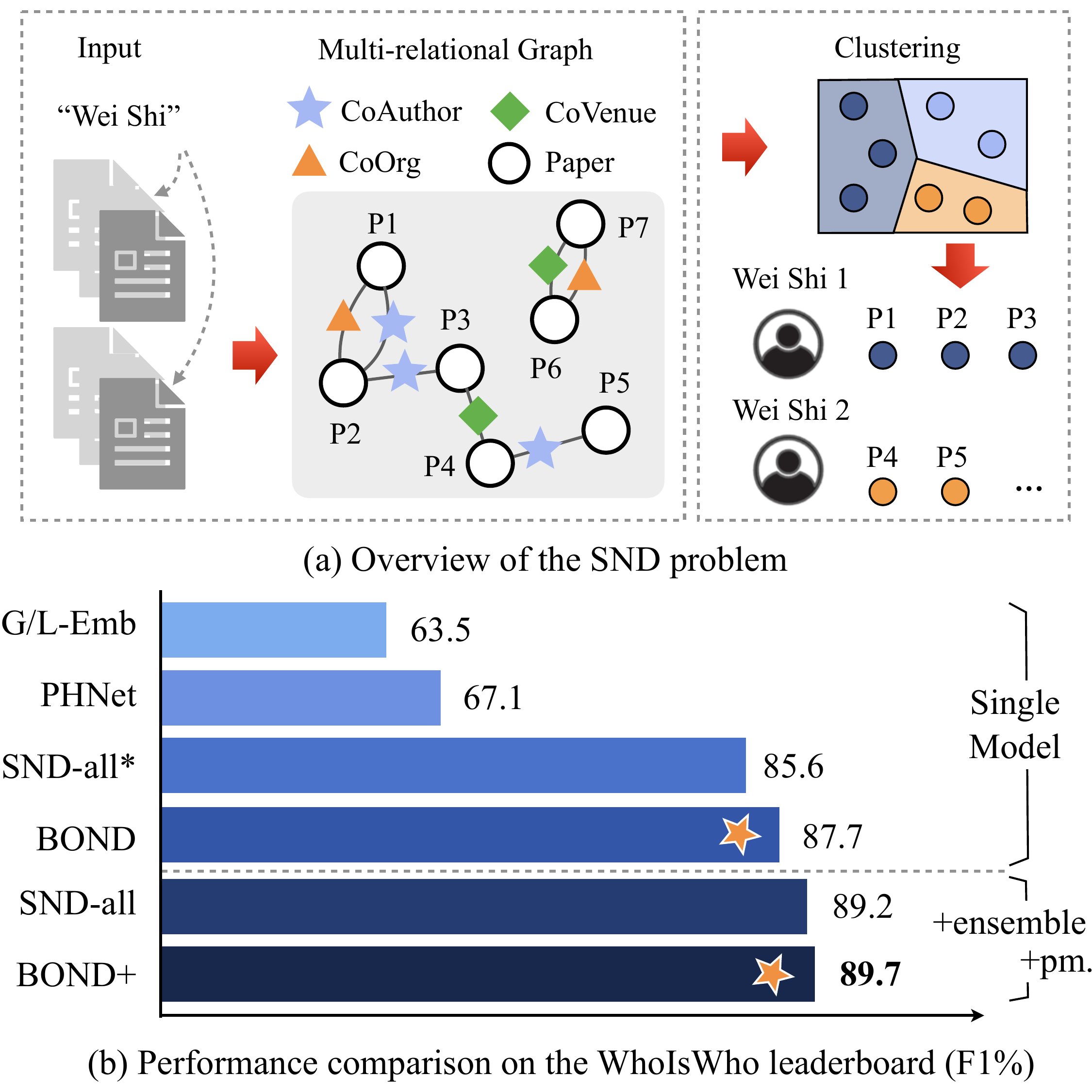}
    \caption{\textbf{
    An overview of the SND problem and performance comparisons between \model and baselines.}
    \textmd{(a) Paper connections are established through diverse relationships. Noise is observed in the linkage of Paper \textit{P4} to Paper \textit{P3}; (b) \textit{SND-all*}: Single Model Version of SND-all, \textit{pm.}: post-match.}}
    \label{fig:task}
\end{figure}

In this paper, we delve into the important task of From-Scratch Name Disambiguation (SND), which is fundamental for building digital libraries. The main goal, as shown in Figure \ref{fig:task} (a), is to organize papers linked to the same author's name into separate author profiles, each representing an individual's work. 
However, due to the missing, fragmented, and noisy paper attributes (e.g. author email, author organizations) across data sources, the performance of SND methods are still unsatisfactory.
Previous research has traditionally treated SND as a clustering problem, which can be broken down into two main tasks:
(1). \textit{Local Metric Learning.} This task concentrates on assessing fine-grained similarities among papers. It typically uses advanced embedding techniques to transform these papers into lower-dimensional representations. Then, metric functions are applied to calculate local pairwise similarities among these papers.
(2). \textit{Global Clustering.} With the learned local relationship of these papers, clustering methods are usually used to acquire the global partition of these papers, where the papers owned by the same author are divided into the same group.  

Unfortunately, previous methods often approached these two stages as two successive decoupled phases. To clarify, early attempts~\cite{louppe2016ethnicity, atarashi2017deep} employed hand-crafted pairwise paper similarity features, in conjunction with traditional classifiers such as SVM~\cite{hearst1998support}, to establish similarity metric functions. Then, during the global clustering phase, algorithms like DBSCAN~\cite{ester1996density} were used to group papers into distinct clusters.
Recent approaches have ventured into building homogeneous paper similarity graphs~\cite{zhang2018name,li2021disambiguating} based on co-author or other relationships, or constructing heterogeneous graphs~\cite{qiao2019unsupervised,santini2022knowledge} to capture high-order connections. 
For example, PHNet~\cite{qiao2019unsupervised} leverages heterogeneous network embedding techniques to obtain paper representations and employs sophisticated clustering methods to categorize papers into clusters.
However, this isolated learning approach faces challenges in effectively combining the information from local pairwise metric learning and global clustering signals. This separation may result in accumulating errors that are difficult to correct during the training process.

\vpara{Present Work.} 
Building upon the insights mentioned above,
we present \model, a \underline{BO}otstrapping From-Scratch \underline{N}ame \underline{D}isambiguation with Multi-task Promoting approach, to bootstrap the local and global informative signals to each other in an end-to-end regime.

Specifically, \model consists of three key components:
1). \textit{Multi-relational Graph Construction}. \model carefully devises strategies for constructing graphs, ensuring the preservation of multi-relational connections among paper nodes.
2). \textit{Local Metric Learning via Edge Reconstruction}. Leveraging a graph auto-encoder with the Graph Attention Network (GAT)~\cite{velivckovic2018graph} as the encoder, \model learns paper representations via edge reconstruction\footnote{Notably, \model can adapt any graph model based on an attention-aggregation scheme as the base encoder.}.
3). \textit{Global Cluster-aware Learning}. \model utilizes DBSCAN, a structural clustering method, for paper clustering. Throughout the training process, global clustering benefits from pseudo-clustering labels derived from the local metric learning module's paper representations. In a reciprocal manner, these global clustering outcomes provide valuable cues for the local metric learning module, resulting in enhanced paper representations. This collaborative interaction substantially improves the quality of the final paper clustering results.

The primary contributions of \model are summarized as follows:
\begin{itemize}[leftmargin=*]
    \item 
    To the best of our knowledge, we are the first to introduce an
    end-to-end bootstrapping strategy for paper similarity learning and paper clustering to address the SND problem.
    \item \model unifies local metric learning and global cluster-aware learning as multi-task promoting, fostering joint learning and mutual enhancement of both modules.
    \item Extensive experimental results highlight substantial performance gains achieved by \model. 
    Notably, even without intricate ensemble and post-match strategies, \model significantly outperforms the previous Top-1 method of WhoIsWho~\cite{chen2023web}. Now, \model currently holds the top position on the WhoIsWho leaderboard$^1$.
\end{itemize}
\section{Related work}
\label{sec:related}

\subsection{Non-graph-based Methods}
\label{subsec: non-graph-base}

Non-graph-based SND methods traditionally rely on the careful definition of hand-crafted features to quantify pairwise paper similarity~\cite{tang2010bibliometric,cen2013author}.
These similarity features are typically classified into two main categories: relational features and semantic features.
Relational features commonly encompass the extraction of coauthor similarity,
which serves as a pivotal signal for distinguishing authors based on their social connections.
On the other hand, semantic similarity features are frequently derived from various attributes such as paper titles, abstracts, keywords, and similar attributes~\cite{louppe2016ethnicity}, aiming to disambiguate authors by assessing the coherence of research topics.
However, these approaches grapple with limitations in their ability to effectively harness the intricate higher-order structure inherent in paper similarity graphs.

\subsection{Graph-based Methods} 

Graph-based SND methods construct either heterogeneous or homogeneous graphs to leverage high-order information~\cite{tang2011unified,shin2014author}. With the development of network representation learning and graph neural networks, some representative methods~\cite{chen2021name,zhang2017name,zhang2019author} have been integrated into the SND problem, enabling the utilization of node features and the graph structure via aggregating information from neighboring nodes. 
In a notable example~\cite{sun2020pairwise}, a heterogeneous graph is employed to model paper connections. A pair-wise RNN network with attention mechanisms is applied for both blocking and clustering. 
Another approach, proposed in~\cite{pooja2022exploiting}, combines two types of graphs: a person-person graph established by connecting papers with shared coauthors and a document-document graph representing the similarity between the content of publications. These methods adhere to the relational and semantic aspects discussed in Section \ref{subsec: non-graph-base}.
However, these approaches 
usually conduct paper similarity learning and clustering separately, thus
facing the challenge of harmonizing local distance metric learning with downstream global clustering tasks. 
In this work, we strive to jointly learn both local and global information within an end-to-end learning framework on multi-relational local linkage graphs.

\subsection{Clustering Methods for SND Problem}

The determination of cluster numbers is a crucial aspect of the SND problem, and it has been the subject of investigation in prior research~\cite{tang2011unified,zhang2018name}. 
Hierarchical clustering algorithms~\cite{mullner2011modern,heller2005bayesian} operate on the premise that papers with higher similarity should be merged initially, followed by the clustering of the resulting merged clusters.
A two-stage algorithm introduced in~\cite{yoshida2010person} leverages the clustering outcomes from the initial stage to generate clustering features for the subsequent stage.
Furthermore, several methodologies have incorporated spectral clustering to enhance the efficiency of clustering procedures~\cite{han2005name,on2012scalable}. Previous work~\cite{xie2016unsupervised} have advanced joint learning by integrating two components, yet they hinges on pre-training the representation model for effective clustering initiation. 

In contrast, our model utilizes DBSCAN as the clustering strategy, which forms clusters based on density and does not necessitate predefined cluster sizes. 
Moreover, we seamlessly integrate the clustering algorithm into our disambiguation framework in an end-to-end manner, facilitating the joint optimization of local metric learning and global clustering.
\section{Problem Definition}

In this section, we present the preliminaries and the problem formulation of from-scratch name disambiguation.

\begin{definition}
\textbf{Paper}. A paper $p$ is associated with multiple attributes, i.e., $p = \{x_1, \cdots, x_F\}$, where $x_f \in p$ denotes the $f$-th attribute
(e.g., co-authors/venues) and 
$F$ is the number of attributes.
\end{definition}

\begin{definition}
\textbf{Author}. An author $a$ contains a paper set, i.e., $a=\{p_1, \cdots, p_{n}\}$, where 
$n$ is the number of papers authored by $a$. 
\end{definition}

\begin{definition}
\textbf{Candidate Papers}. Given a name denoted by $na$, $\mathcal{P}^{na} = \{p^{na}_1, \dots, p^{na}_N\}$ is a set of candidate papers authored by individuals with the name $na$.
\end{definition}

\begin{problem}
\textbf{From-scratch Name Disambiguation (SND)}. Given candidate papers $\mathcal{P}^{na}$ associated with name $na$,
SND aims at finding a function $\Phi$ to partition $\mathcal{P}^{na}$ into a set of disjoint clusters $C^{na}$, i.e.,
	
\beq{
		\Phi(\mathcal{P}^{na}) \rightarrow C^{na}, \text{where } C^{na}=\{C^{na}_1, C^{na}_2, \cdots, C^{na}_K\}, \nonumber
}

\noindent where $C^{na}$ represents the resulting clusters, each cluster consists of papers from the same author, i.e., $\mathbb{I}(p_i^{na})=\mathbb{I}(p_j^{na}), \forall (p_i^{na}, p_j^{na}) \in C_k^{na}\times C_k^{na}$,  and different clusters 
contain papers from different authors, i.e., $\mathbb{I}(p_i^{na})\ne \mathbb{I}(p_j^{na}), \forall  (p_i^{na}, p_j^{na}) \in C_k^{na}\times C_{k'}^{na}, k \ne k'$. $\mathbb{I}(p_i^{na})$ is the author identification of the paper $p_i^{na}$.
\end{problem}

Notably, \model tries to tackle the SND problem based on the built paper-author multi-relational graphs (see Section~\ref{subsec: graphcons} for detailed information). Compared to the traditional methods which are based on non-graph-based methods. Recent attempts~\cite{qiao2019unsupervised,sun2020pairwise} imply that building relational graphs can characterize the fine-grained correlations among papers and authors, thus facilitating the following SND algorithms. The experimental results also indicate the graph-based SND framework consistently outperforms other non-graph-based ones ranging from 6.0\% to 32.6\%.

\section{Methodology}

\begin{figure*}
    \centering
    \includegraphics[width=17cm]{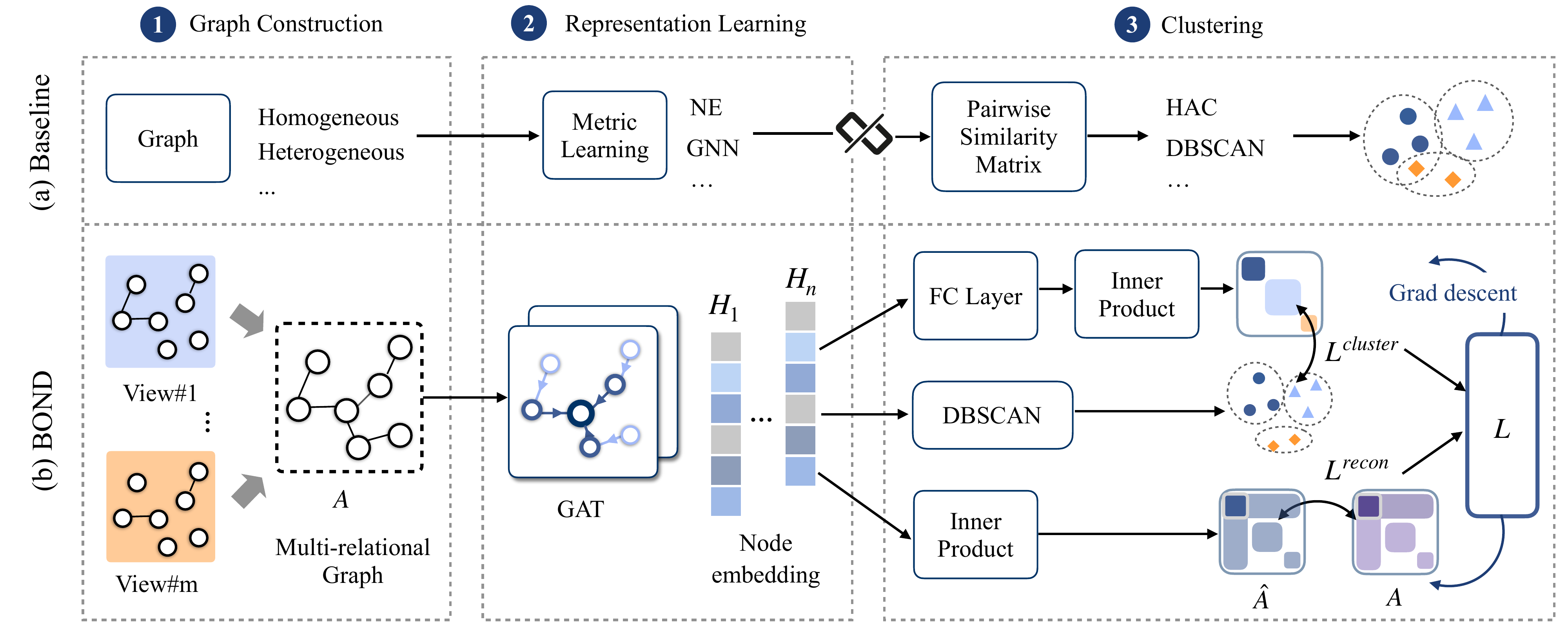}
    \caption{\textbf{The overall framework of \model and other SND methods.} \textmd{NE: network embedding; HAC: hierarchical agglomerative clustering; our proposed framework, as depicted in (b), integrates metric learning and clustering within a multi-task learning framework. By optimizing the weighted sum of reconstruction loss and cluster-aware loss, the global information derived from the clustering component can reciprocally guide the local information extracted from the reconstruction part.}}
    \label{fig:model1}
\end{figure*}

As previously discussed, conventional approaches 
typically adopt a decoupled pipeline for addressing the from-scratch name disambiguation problem. This pipeline involves initially capturing local relationships among papers and subsequently performing global clustering based on the localized information. Regrettably, this two-stage optimization process hinders the seamless diffusion of information between the two distinct task modalities, making it challenging to self-correct cumulative errors.
In response to this limitation, we introduce \model, an end-to-end approach for name disambiguation. 
It starts by building a multi-relational graph to capture paper relationships (Section~\ref{subsec: graphcons}). 
Then, local metric learning is performed to enhance paper representations (Section~\ref{subsec: locallinkage}), and a clustering-aware learning algorithm is used to understand global relationships (Section~\ref{subsec: globalalign}). 
Finally, \model optimizes both tasks together within an end-to-end algorithm (Section~\ref{subsec: jointopt}). 
The framework is illustrated in Figure \ref{fig:model1}. In the following sections, we delve into the specifics of each individual component.

\subsection{Multi-relational Graph Construction}
\label{subsec: graphcons}

To estimate local relationships, i.e., pairwise similarities, among candidate papers, 
we create a local linkage graph, denoted as $G^{na} = (\mathcal{P}^{na}, E^{na})$, for each name $na$. 
Here, $\mathcal{P}^{na}$ is the set of candidate papers, 
and $E^{na} \in \mathcal{P}^{na} \times \mathcal{P}^{na}$ represents the edge set between these papers. 
To ensure the preservation of comprehensive relationships while eliminating extraneous connections among papers, 
it is imperative to precisely specify the edges and node features.

\vpara{Edge Construction.}
We measure paper similarities through multiple pathways that signify authorship, 
such as co-author (authored by individuals with the same name, except for the disambiguated name), 
co-venue (sharing the same conference or journal), 
and co-organization (affiliated with the same institution). 
While traditional approaches~\cite{fan2011graph, pooja2021exploiting} have frequently relied on the co-author relationship as a primary measure of paper similarities, recent empirical research~\cite{chen2023web} has shed light on the effectiveness of alternative paper attributes in capturing semantic or structural aspects of paper similarity. 
In light of these findings, we opt to incorporate three distinct paper attributes—namely, co-author, co-org, and co-venue—as factors for measuring paper similarity.

We employ different linguistic word-match metrics to capture the exact and relative similarities between these paper attributes. For co-author and co-venue relationships, we use the \textit{word overlap} metric to calculate similarities between papers. 
However, for co-organization relationships, where the attribute often contains redundant words, we use the \textit{Jaccard Index} as the metric. We determine whether to add edges between papers based on thresholds determined through validation performance. Our experiments in Section~\ref{subsec: ensemthresh} indicate that the performance is sensitive to these pre-defined thresholds.

\vpara{Node Feature Initialization.}
The semantic information captured by node input features is equally essential for identifying paper authorship. 
Following the analysis in ~\cite{chen2023web}, the combination of paper titles, author organizations, and keywords proves to be crucial, thus we also adopt these paper attributes to initialize the input features.
For simplicity and effectiveness, we train a Word2Vec~\cite{mikolov2013efficient} model on the \da corpus and encode each word in the relevant paper attributes into a low-dimensional continuous vector. The superiority of Word2Vec is discussed in Section \ref{subsec: pretrain}. These vectors are then summed to create paper embeddings $X_i$.

\subsection{Local Metric Learning}
\label{subsec: locallinkage}

In the absence of supervised authorship signals within the candidate papers, 
we rely on semantic and structural paper features for quantifying paper similarities. 
Existing approaches often employ unsupervised paper embeddings obtained through network embedding (NE)~\cite{zhang2017name,qiao2019unsupervised} or graph neural networks (GNNs)~\cite{qiao2019unsupervised,sun2020pairwise}. 
Similarly, we employ a graph auto-encoder~\cite{kipf2016variational}, comprising an encoder and a decoder, for the purpose of learning precise paper representations. 
The encoder leverages GAT due to their adaptability in learning edge weights through the attention mechanism.
The paper representations are derived through the following expression,

\beq{
\label{eq:linkage_h}
    H^{''} = \text{GAT}(W_e, A(\mathcal{G}), H^{'}) = g(A(\mathcal{G}) H^{'} W_e^{\top} + b_e) ,
}

\noindent where $H^{'}$ represents the input paper embeddings (set to $X$ in the first layer),
while $W_e$ and $b_e$ denote the projection matrix and the bias of the encoder, respectively.
$A(\mathcal{G})$ represents the learned attention matrix,
and $g$ is the activation function.
The edge weight is parameterized as follows,
\begin{equation}
    e_{ij} = c^{\top} ([W_e H_i^{'}||W_e H_j^{'}]), j \in N_i ,
\end{equation}
For node $i$, we calculate the coefficients between $i$ and its neighbors $j$ separately. $W_e$ is a shared parameter matrix to extend dimension and $c$ is for projecting the high-dimension to a real number.
$||$ is the concatenation operator.
The attention weight in $A(\mathcal{G})$ is calculated as follows, 
\begin{equation}
    \alpha_{ij} = \frac{\text{exp}(e_{ij})}{\sum_{k\in N_i} \text{exp}(e_{ik})} ,
\end{equation}

We utilize multi-head attention to obtain richer hidden representations and employ two GAT layers in the encoder to obtain hidden embeddings $H$.

The decoder is defined as the inner product between the hidden embeddings,
\begin{equation}
\label{eq:recon_innerp}
    \hat{A} = \text{sigmoid}(H^{\top} H) .
\end{equation}

The objective function is designed to minimize the reconstruction error of the adjacency matrix through the cross-entropy loss,

\begin{equation}
    \label{eq:recon_loss}
    \mathcal{L}^{\text{recon}} = 
    \sum_{i=1}^N\sum_{j=1}^N (A_{ij} \log p (\hat{A}_{ij}) + (1- A_{ij}) \log (1 - p (\hat{A}_{ij}))) ,
\end{equation}

\noindent where $A$ is the original adjacency matrix of $\mathcal{G}$
 and $N$ is the node number in the graph. The local minima achieved through reconstructing the linkage among papers yields appropriate paper representations, forming the foundation for the subsequent process.

\subsection{Global Cluster-aware Learning}
\label{subsec: globalalign}

In traditional methodologies, the paper embeddings denoted as $H$, which result from local linkage learning, are typically employed for estimating pairwise similarities between papers. 
Subsequently, these methods utilize clustering algorithms like DBSCAN to partition the papers into distinct clusters to achieve disambiguation. However, a common oversight in these approaches is the underutilization of global clustering results, which have the potential to enhance the quality of paper representations obtained through local optimization. 
We posit that it is possible to effectively perform paper similarity learning and clustering in an end-to-end manner, thereby capitalizing on the mutual reinforcement of these two tasks.

To this end, we leverage DBSCAN to generate cluster labels due to its flexibility in cluster number specification, denoted as $Y$, based on the paper embeddings $H$. 
These labels provide essential global alignment signals. 
To capitalize on these signals and enhance the quality of paper representations, we introduce a fully connected layer, which processes the paper embeddings $H$ to produce output representations $C$, aiming to learn cluster-aware representations,

\beq{
\label{eq:cluster_c}
C = H W_c^{\top} + b_c ,
}

\noindent where $W_c$ and $b_c$ represent the projection matrix and bias parameters of the fully connected layer, respectively.

Then, we attain the pairwise relationships $\mathcal{C}$ between nodes through inner product operations, i.e., $\mathcal{C} = C C^{\top} $. To facilitate a comparison between the global alignment label $Y$ generated by DBSCAN and the local results $\mathcal{C}$, we also convert $Y$ into the adjacency matrix $\mathcal{Y}$,
\beq{
\mathcal{Y} = [\mathbb{I} (Y_i = Y_j)]^{N \times N} ,
}

\noindent where $\mathcal{C}_{ij}$ indicates the similarity score between node $i$ and node $j$,
while $\mathcal{Y}_{ij}$ signifies whether node $i$ and node $j$ belong to the same cluster label\footnote {Here we regard nodes with label $-1$ as the same cluster for simplicity.}.

Finally, we define the cluster-aware loss using the cross-entropy objective to bootstrap the global alignment signals to the local linkage learning module,
\beq{
\label{eq:cluster_loss}
\mathcal{L}^{\text{cluster}} = 
\sum_{i=1}^N\sum_{j=1}^N (\mathcal{Y}_{ij} \log p (\mathcal{C}_{ij}) + (1- \mathcal{Y}_{ij}) \log (1 - p (\mathcal{C}_{ij}))) .
}

\subsection{Joint Objective Optimization}
\label{subsec: jointopt}

In this process, we aim to find a balance between the cluster-aware loss $\mathcal{L}^{\text{cluster}}$ and the reconstruction loss $\mathcal{L}^{\text{recon}}$, which are crucial components for our \model. We achieve this by using a weighted sum of these losses, as represented by the following equation:

\begin{equation}
    \mathcal{L} = \lambda \mathcal{L}^{\text{cluster}} + (1-\lambda) \mathcal{L}^{\text{recon}}
\end{equation}
where $\lambda$ is a hyper-parameter empirically set to $0.5$. 
We employ the clustering labels $Y$ of the last epoch as the final prediction results.

The training procedure of \model is outlined in Algorithm \ref{algo:joint}. 
For each epoch,
in line $2$-$3$, we obtain hidden representation $H$ via GNN encoders and cluster-aware representation $C$ successively.
In line $4$, we get the outputs $\hat{A}$ and $\mathcal{C}$ of local metric learning and cluster-aware learning, respectively.
In line $5$, pseudo labels $Y$ are generated based on hidden representation $H$.
Finally, in line $6$-$8$, we compute the total loss $\mathcal{L}$ based on separate loss of each task and then optimize the model via back propogation.

Local metric learning serves the purpose of enhancing the model's comprehension of paper similarities and the underlying graph topology. 
However, it has a vulnerability to noise, which may stem from local linkage graphs constructed based on feature similarity.
In contrast, global cluster-aware learning aligns representations with the goal of the SND problem. These two tasks offer diverse perspectives and mutually enhance each other.

\begin{algorithm}
    \SetKwFunction{innperProduct}{innperProduct}
    \SetKwInOut{KwIn}{Input}
    \SetKwInOut{KwOut}{Output}

    \KwIn{Multi-relational Graph $G^{na}$, the multi-task loss $\mathcal{L}^{\text{cluster}}$, $\mathcal{L}^{\text{recon}}$ and the loss weight $\lambda$. (GD: gradient descent).}
    \KwOut{Obtain model with parameters $\theta$.}
    \For{$\text{iter}=1,2,\cdots,T$}{
        Get hidden representation $H$ with Eq.(\ref{eq:linkage_h}) via local metric learning.\\
        Get cluster-aware representation $C$ with Eq.(\ref{eq:cluster_c}) on $H$.\\
        Get reconstruction adjacency matrix $\hat{A}$ with Eq.(\ref{eq:recon_innerp}) and pairwise class proximity matrix $\mathcal{C}$. \\ 
        Get pseudo-label $Y$ with DBSCAN on $H$.\\
        Compute reconstruction loss $\mathcal{L}^{\text{recon}}$ with Eq.(\ref{eq:recon_loss}) and cluster-aware loss $\mathcal{L}^{\text{cluster}}$ with Eq.(\ref{eq:cluster_loss}).\\
        Calculate the joint loss $\mathcal{L}$ as the weighted sum of $\mathcal{L}^{\text{recon}}$ and $\mathcal{L}^{\text{cluster}}$.\\
        Update $\theta$ via GD on $\nabla_\theta  \mathcal{L}$.
    }
    \caption{The Joint Objective Optimization Procedure}
    \label{algo:joint}
\end{algorithm}

\subsection{Time Complexity}
The local metric learning module adopts GAT, thus the time complexity of layer $k$ is $\mathcal{O}\left(D_k^2N+D_kE\right)$, where $D_k$ is the embedding size in layer $k$, $N$ is the number of nodes and $E$ is the number of edges. 
The global clustering module adopts DBSCAN whose average time complexity is $\mathcal{O}\left(N\log N\right)$.
The time complexity to build the reconstruction adjacency matrix is $\mathcal{O}\left(N^2D_k\right)$.
Since the embedding size is far smaller than the number of nodes or edges, the time complexity of \model is $\mathcal{O}\left(N^2+E\right)$.

\section{Experiments}
The source code for this work is openly accessible to the public\footnote{https://github.com/THUDM/WhoIsWho}.

\subsection{Experimental Setup}

\vpara{Datasets.}
We utilize the \da-v3 dataset~\cite{chen2023web} as our experimental benchmark, which is the largest human-annotated name disambiguation dataset to date. This dataset comprises 480 unique author names, 12,431 authors, and 285,252 papers, each with attributes like title, keywords, abstract, authors, affiliations, venue, and publication year. Following the \da competition, we divide it into training, validation, and testing sets in a $2:1:1$ ratio based on author names.

\vpara{Baselines.}
We've conducted a rigorous comparison of our method with various SND approaches. 
To ensure fairness, the number of clusters has been aligned with the true value.

\begin{itemize}[leftmargin=*]
    \item \textbf{Louppe et al. ~\cite{louppe2016ethnicity}}: 
    employs a classification model trained for each paper pair, aiming to determine if they are authored by the same individual. 
    They utilize carefully designed features and semi-supervised cut-off strategies to form flat clusters of papers.
    \item \textbf{IUAD ~\cite{li2021disambiguating}}: constructs paper similarity graphs based on co-author relationships. It enhances the collaboration network using a probabilistic generative model that integrates network structures, research interests, and research communities.

    \item \textbf{G/L-Emb ~\cite{zhang2018name}}:
    utilizes common features between papers to create paper-paper networks. It learns paper representations by reconstructing these networks and employs hierarchical agglomerative clustering (HAC) for clustering.

    \item \textbf{LAND ~\cite{santini2022knowledge}}: 
    constructs a knowledge graph with papers, authors, and organizations as nodes and multi-relational edges. It uses BERT~\cite{devlin2018bert} for initializing entity embeddings and employs the LiteralE~\cite{kristiadi2019incorporating} knowledge representation learning method. Then, it also uses HAC for clustering.
 
    \item \textbf{PHNet ~\cite{qiao2019unsupervised}}: 
    builds a heterogeneous paper network and employs heterogeneous graph convolution networks (HGCN) for node embeddings. It uses graph-enhanced HAC for clustering, requiring a predefined cluster size.

    \item \textbf{SND-all ~\cite{chen2023web}}: 
    applies metapath2vec for extracting heterogeneous relational graph features along with soft semantic features. It utilizes DBSCAN for clustering and involves bagging in network embedding training. Additionally, it employs a rule-based post-match algorithm for handling outliers and cluster formation.

\end{itemize}

\vpara{Evaluation Metric}
The evaluation of clustering results is based on pairwise Precision, Recall, and F1~\cite{chen2023web, zhang2018name}. 
Subsequently, a macro metric is derived by averaging these performance metrics across all the individual names.

\subsection{Main Results}

In Table \ref{tb: baseline_pf}, we conduct a comprehensive comparative analysis of various author disambiguation methods. 
Louppe et al. distinguishes itself by relying on supervised pair-wise classification, underpinned by meticulously designed features. 
In contrast, other methodologies adopt unsupervised techniques for learning from raw data.

As an illustration, IUAD establishes coauthor networks through the mining of frequent collaborative relationships, subsequently incorporating probabilistic generative models that leverage similarity functions within the collaborative network. 
The relatively suboptimal performance of IUAD can be attributed to its heavy reliance on co-author relationships. 
In contrast, our approach considers a broader spectrum of relationships, thereby preserving comprehensive structural paper connections.
Furthermore, when juxtaposed with G/L-Emb, LAND, PHNet and SND-all, each of which takes into account distinct types of connections, our model emerges as a notable frontrunner in terms of performance.
G/L-Emb enhances local distance learning between papers through global semantic representations. LAND leverages knowledge embedding, while PHNet harnesses the capabilities of a heterogeneous graph neural network, and SND-all deftly integrates soft semantic features with heterogeneous relational graph features. 
Notably, our approach stands apart by operating as an end-to-end solution for author disambiguation, seamlessly harmonizing the twin processes of learning paper similarities and conducting clustering. 
This harmonious integration culminates in the generation of remarkably discriminative representations, thereby distinguishing our methodology from the decoupled approaches of our counterparts.

\begin{table}
	\newcolumntype{?}{!{\vrule width 1pt}}
	\newcolumntype{C}{>{\centering\arraybackslash}p{4em}}
	\caption{
		\label{tb: baseline_pf} \textbf{Results of from-scratch name disambiguation (\%).} 
	}
	\centering 
	\renewcommand\arraystretch{1.0}
	\begin{tabular}{@{~}l?@{~}*{1}{CCC}@{~}}
		\toprule
		\textbf{Models} & \textbf{Precision} & \textbf{Recall} & \textbf{F1} \\
		\midrule
            Louppe et al. & 68.05              & 46.32           & 55.12       \\
            IUAD          & 58.82              & 65.22           & 61.63       \\
		G/L-Emb       & 50.77              & 84.64           & 63.48       \\
            LAND          & 61.20              & 61.12           & 61.12       \\
            PHNet         & 65.91              & 68.32           & 67.09       \\
            SND-all*      & 81.68	           & 89.97	         & 85.62       \\
		\midrule
		\model        & \textbf{82.07}     & \textbf{94.21}  & \textbf{87.72}  \\
		
		\bottomrule
	\end{tabular}
\end{table}

\begin{table}[ht]
	\newcolumntype{?}{!{\vrule width 1pt}}
	\newcolumntype{C}{>{\centering\arraybackslash}p{4em}}
	\caption{
		\label{tb: multi-loss}\textbf{Improvement of unified loss (\%) and the statistical significance.} \textmd{\textit{Only Cluster.}: training only on Cluster-aware Loss; \textit{Only Recon.} training only on Local Metric Learning Loss;}
	}
	\centering 
	\renewcommand\arraystretch{1.0}
	\begin{tabular}{@{~}l?@{~}*{1}{CCCC}@{~}}
		\toprule
		\textbf{Loss} & \textbf{Precision} & \textbf{Recall} & \textbf{F1} & \textbf{P-value}\\
		\midrule
		
            Only Cluster.   & 79.83              & \textbf{96.42}    & 87.35    & 0.0014   \\
            Only Recon.      & 77.58              & 94.19           & 85.08     & 9.4640E-5  \\
		\midrule
		Unified loss   & \textbf{82.34}     & 95.27  & \textbf{88.33} & / \\
		
		\bottomrule
	\end{tabular}
\end{table}

\subsection{Ablation Study}

In this section, we provide a justification for the effectiveness of each component within our framework.

\vpara{Effect of the different losses.}
As depicted in Table \ref{tb: multi-loss}, we 
compare the performance of joint loss, i.e., $\mathcal{L}$, and the use of single loss, i.e., $\mathcal{L}^\text{recon}$ and $\mathcal{L}^\text{cluster}$
, on the validation set.
The performance of the cluster-aware learning task surpasses the local metric learning task, suggesting that downstream clustering tasks can provide more 
accurate
guidance for representation learning. 
The unified task demonstrates an improvement of $+0.98\%$ over the cluster-aware learning task and $+3.25\%$ compared to the local metric learning task. 
These findings validate the efficacy of unifying the two tasks,
as they complement and enhance one another. 

\begin{figure}[t]
	\centering
	\subfigure[Convergence of different losses]{\label{subfig:jointrain}
		\includegraphics[width=0.23\textwidth]{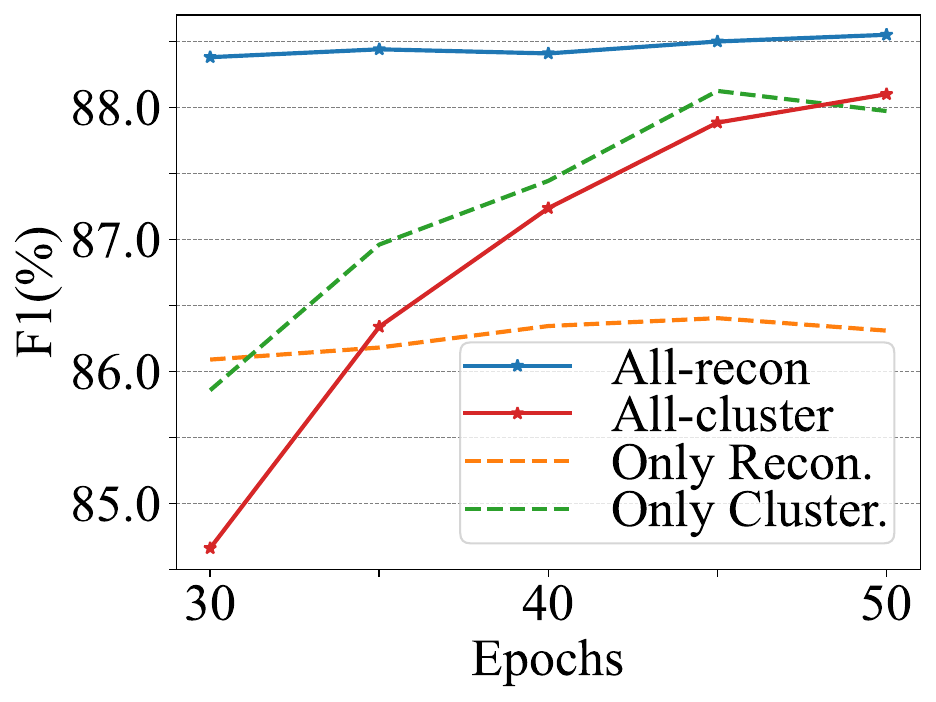}
	}
	\hspace{-0.1in}
	\subfigure[Local metric learning only]{\label{subfig:recontrain}
		\includegraphics[width=0.23\textwidth]{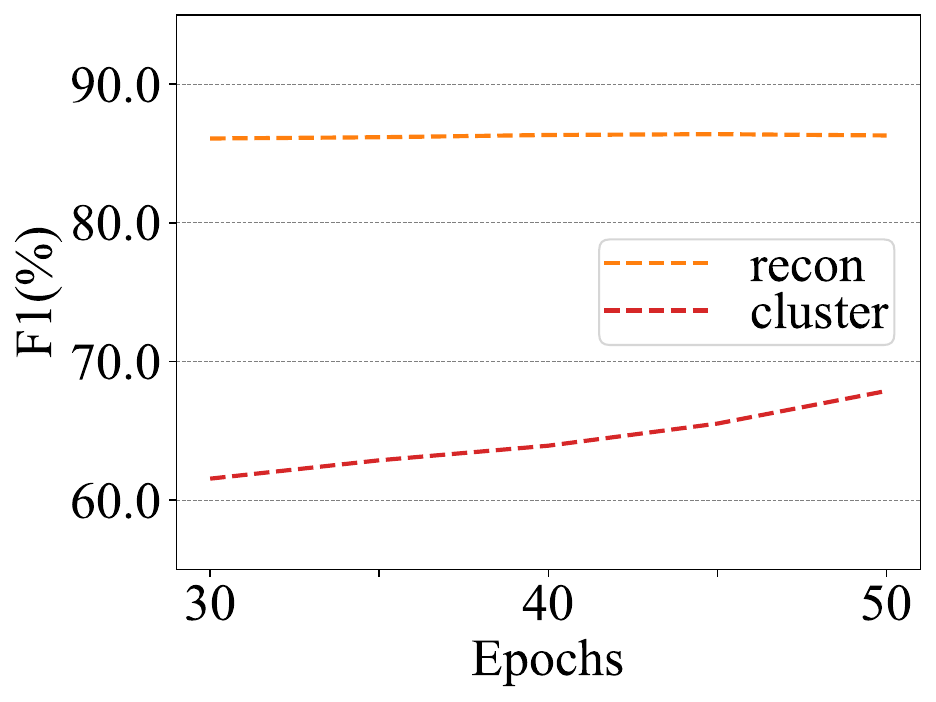}
	}
	\hspace{-0.1in}
	\subfigure[Multi-relational features]{\label{subfig:multirel}
		\includegraphics[width=0.23\textwidth]{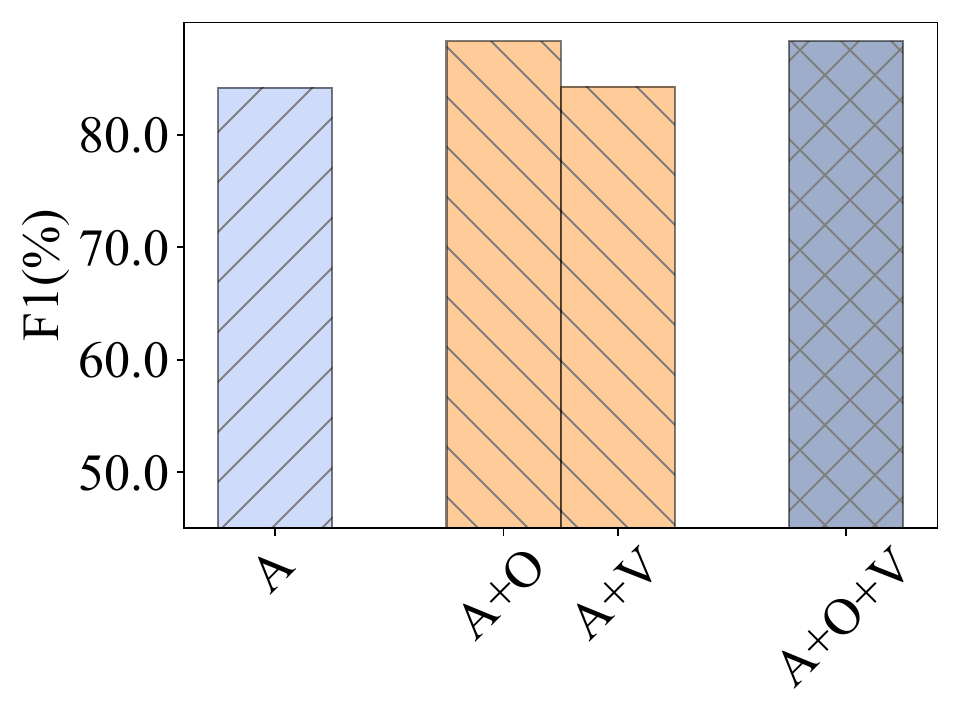}
	}
	\hspace{-0.1in}
	\subfigure[Different name performances]{\label{subfig:multiname}
		\includegraphics[width=0.23\textwidth]{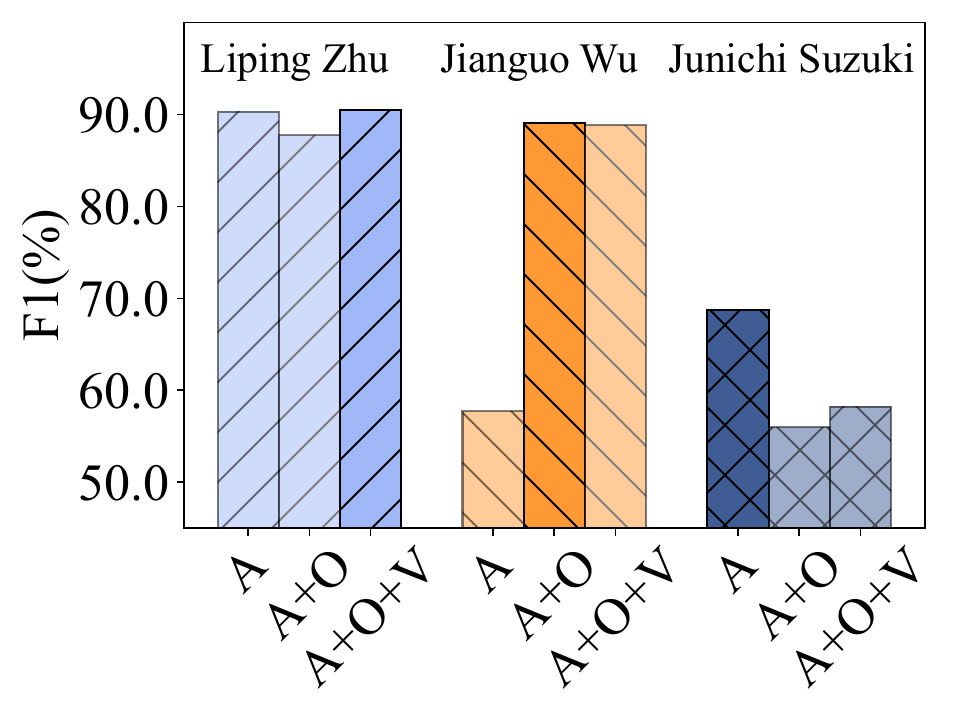}
	}
	\vspace{-5pt}
	\caption{\label{fig:feature_import} \textbf{Effect of different losses and multi-relational features.}\textmd{(a): \textit{All}: training on all loss; \textit{All-recon}: the clusters of local metric learning; \textit{All-cluster}: the outputs of cluster-aware learning. (b): Training only on Local Metric Learning Loss. \textit{recon}: the clusters of local metric learning; \textit{cluster}: the outputs of cluster-aware learning. (c) and (d): \textit{A}: CoAuthor; \textit{O}: CoOrg; \textit{V}: CoVenue.}}
\end{figure}

We further compare the performance achieved by the unified loss with the single loss as the training goes, as illustrated in Figure \ref{subfig:jointrain}. 
The reconstruction performance of the unified loss, i.e., the blue line, is better than the results with the model using single reconstruction loss for training, i.e., the orange line. 
While in Figure \ref{subfig:recontrain}, take single reconstruction loss as an example, training processes in a two-stage way, 
causing a disconnect between metric learning and clustering. 
The results indicate that joint optimization can enhance the performance of both tasks, achieving superior results compared to separate single-task approaches.

\vpara{Effect of multi-relational features.}
Figure \ref{subfig:multirel} demonstrates the impact of multi-relational features. 
Our study of multi-view graphs is constructed in a cumulative fashion. 
CoA denotes the co-author relationships, excluding the author to be disambiguated, and it yields high-quality relations.
CoO represents co-organization relationships of the disambiguation author.
CoV refers to the co-venue relationships of the compared two papers.

The performance of Co(A+O) surpasses that of CoA by $+4.13\%$, suggesting that co-organization contains valuable information and fills the gap that co-author cannot cover. 
Since co-venue relationships are not that discriminative to represent the authorship,
we set the probability to 0.1 to reserve the co-venue edges. 
In our study, CoV doesn't take effect for the single model of \model,
but achieves clear improvements for our ensemble model when combined with CoA and CoO relational features. 

However, Figure \ref{subfig:multiname} substantiates the distinct characteristics of local linkage graphs across different names by manipulating the multi-relational graphs employed by \model. 
For example, in the case of Jianguo Wu, the incorporation of the Co-organization relation\footnote{corresponding to the threshold in Figure \ref{fig: ensemble}} results in a performance improvement of +31.37\%. This finding suggests that Co-organization uncovers information that is not present in Co-author relationships. In contrast, the performance of the names Liping Zhu and Junichi Suzuki is compromised, indicating that Co-organization may introduce noise in these instances. Similarly, the Co-venue enhances performance by +2.73\% in Liping Zhu and +2.23\% in Junichi Suzuki. However, it either weakens or has no effect on Jianguo Wu. These results imply that the Co-author and Co-organization relationships already provide sufficient information for disambiguating these author names.

In light of these observations, our motivation is directed towards the ensemble of diverse models by employing edge-purging strategies. This approach will be elucidated in the following section.

\subsection{\da Competition}
\label{subsec: ensemthresh}
To assess the effectiveness of our proposed approach, we have extended the \model to be evaluated on the widely recognized \da benchmark\footnote{http://whoiswho.biendata.xyz/}, which has attracted the attention of over 3,000 researchers. Notably, our model, bolstered by ensemble learning techniques and the introduction of a post-match strategy (denoted as BOND+), has remarkably secured the first position in this benchmark.
In the following subsection, we provide a detailed introduction to these enhanced strategies and conduct a meticulous ablation analysis to comprehensively evaluate their influence.

\begin{figure}[ht]
    \centering
    \includegraphics[width=0.48\textwidth]{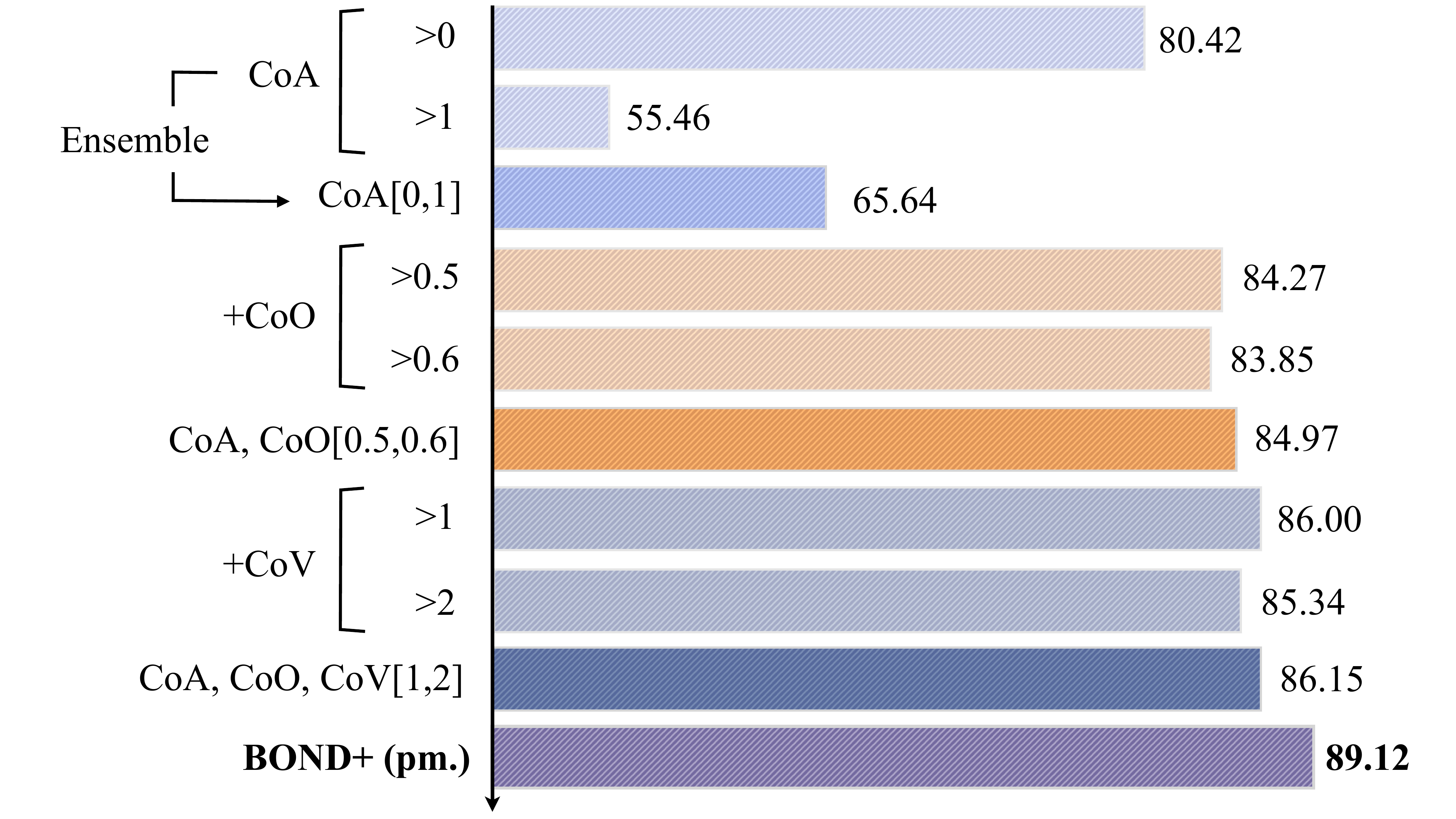}
    \caption{\textbf{The results of ablation analysis of our ensemble model. (\%)} \textmd{">0" signifies that the threshold is set to 0. "CoA[0,1]" denotes the ensembling of models with coauthor values greater than 0 and 1, respectively. "CoA, CoO[0.5,0.6]" refers to the multi-view model with coauthor and coorg edges, when coauthor greater than 0 and 1 and coorg greater than 0.5 and 0.6, respectively.
    }}
    \label{fig: ensemble}
\end{figure}

\vpara{Ensemble learning.}
Since different multi-relational features provide different inductive biases for name disambiguation,
we argue that the ensemble of multiple models trained on local linkage graphs built with different relational features could complement each other.
In this study, we train multiple models with different relational features and employ a voting mechanism for their output labels. 
As illustrated in Figure \ref{fig: ensemble}, 
 an increase in the number of models 
can result in a performance enhancement of up to $+5.73\%$.

\vpara{Post-match.}
Outliers generated by DBSCAN can be post-matched to either existing paper clusters or new clusters. 
Following the idea of \da contest winners,
we conduct similarity matching between unassigned papers (outliers) and assigned papers based on paper titles, keywords, co-authors, co-venues (CoV), and co-organizations (CoO).
We adopt \textit{tanimoto distance} to calculate CoO and CoV similarities
and character matching on paper keywords and titles.
If the combined similarity score exceeds a pre-defined threshold, i.e., 1.5 in our method, the papers are assigned to their respective groups.
As illustrated in Table \ref{fig: ensemble}
, post-match improves the performance by +2.97\%. 

\subsection{Transductive v.s. Inductive Learning}

In this section, we scrutinize the performance of our model in both transductive and inductive scenarios.
In the transductive context, we pursue the training of distinct models for each graph, which is constructed for individual names. 
Consequently, we adjust the dimensions of the output representations $C$ in accordance with the specific number of nodes within the given graph.

In the inductive setting, we train the model using all graphs in the training set, which are randomly shuffled in each epoch. The size of the fully connected layer $C$ is fixed. Subsequently, the model is frozen during inference on unseen graphs in the test set. 

As depicted in Table \ref{tb: tran_ind}, the transductive setting exhibits a performance improvement of +2.36\% compared to the inductive setting,
also with an absolute gain of 1.04\% over a fixed size of $C$, 
indicating that the transductive setting with adaptive output size suits SND problem most.
This superiority can be attributed to the transductive approach's capability to capture the unique characteristics of each graph pertaining to individual names. 
Additionally, the adaptability of the fully connected layers, accommodating different graph sizes, contributes to the observed performance gain.

\begin{table}
	\newcolumntype{?}{!{\vrule width 1pt}}
	\newcolumntype{C}{>{\centering\arraybackslash}p{4em}}
	\caption{
		\label{tb: tran_ind}Transductive learning and inductive learning (\%)
	}

	\centering 
	\renewcommand\arraystretch{1.0}
	\begin{tabular}{@{~}l?@{~}*{1}{CCC}@{~}}
		\toprule
		\textbf{Settings} & \textbf{Precision} & \textbf{Recall} & \textbf{F1} \\
		\midrule
            Transductive            & \textbf{85.18}    &   94.97   &   \textbf{88.55}   \\
            
            Transductive-fixed      & 83.24     &   \textbf{95.65}   &  87.51    \\
            Inductive               & 84.15     &   91.49   & 86.19     \\
		\bottomrule
	\end{tabular}
\end{table}

\subsection{Can Pre-trained Models help?}
\label{subsec: pretrain}

\begin{table}
	\newcolumntype{?}{!{\vrule width 1pt}}
	\newcolumntype{C}{>{\centering\arraybackslash}p{4em}}
	\caption{
		\label{tb:emb_model}Semantic embedding methods (\%).
	}
	\centering 
	\renewcommand\arraystretch{1.0}
	\begin{tabular}{@{~}l?@{~}*{1}{CCC}@{~}}
		\toprule
		\textbf{Methods} & \textbf{Precision} & \textbf{Recall} & \textbf{F1} \\
		\midrule
            OAG-BERT & \textbf{82.39} &   91.58   &   86.74\\
            SciBERT & 76.64 &   95.15   &   84.90 \\
            Word2vec &82.36	&   \textbf{95.25}	&   \textbf{88.34} \\
		\bottomrule
	\end{tabular}
\end{table}

For GNN encoders,
we employ Word2Vec to initialize node features.
We conduct a comparative analysis between Word2Vec and other pre-trained models, including OAG-BERT ~\cite{liu2022oag} and SciBERT ~\cite{Beltagy2019SciBERT}. OAG-BERT is pre-trained on the corpus of Open Academic Graph~\cite{zhang2019oag}, while SciBERT is trained based on papers in the Semantic Scholar corpus. 
We use the \texttt{oagbert-v2-sim} version of OAG-BERT,
which is fine-tuned on \da training corpus.
As illustrated in Table \ref{tb:emb_model}, Word2Vec surpasses OAG-BERT by 1.84\% and outperforms SciBERT by 4.05\%,
showing the clear gap between semantic knowledge embodied in large pre-trained models 
and the discriminative information required by name disambiguation task.


This observation suggests that large pre-trained models may embody substantial semantic knowledge from extensive datasets, but they exhibit noticeable bias when compared to the discriminative information required for the name disambiguation task.

\subsection{Clustering Robustness}

\begin{table}
	\newcolumntype{?}{!{\vrule width 1pt}}
	\newcolumntype{C}{>{\centering\arraybackslash}p{4em}}
	\caption{
		\label{tb: cluster}Clustering methods (\%).
	}
	\centering 
	\renewcommand\arraystretch{1.0}
	\begin{tabular}{@{~}l?@{~}*{1}{CCC}@{~}}
		\toprule
		\textbf{Methods} & \textbf{Precision} & \textbf{Recall} & \textbf{F1} \\
		\midrule
            DBSCAN & \textbf{82.36}	&95.25	&\textbf{88.34}\\
            HDBSCAN & 81.94	&\textbf{95.52}	&88.21 \\
            AP Clustering & 70.21	&72.78	&71.47\\
            OPTICS &82.19	&95.27	&88.25 \\
		\bottomrule
	\end{tabular}
\end{table}

Noise significantly impacts early-stage clustering in name disambiguation due to the random initialization of representation learning models, which produce initial low-quality embeddings. To address this, our approach emphasizes the selective propagation of high-confidence labels, sidelining those of low confidence and high noise. We implement a denoising module, leveraging DBSCAN's mechanism to distinguish and exclude unclear node labels, thus prioritizing clear, high-confidence labels for training. 

Furthermore, as indicated in Table ~\ref{tb: cluster}, similar denoising capabilities can be found in clustering algorithms like HDBSCAN~\cite{mcinnes2017accelerated} and OPTICS~\cite{ankerst1999optics}, contrasting with AP Clustering~\cite{frey2007clustering}, which lacks a denoising process and thereby introduces noise. Our findings demonstrate that our framework can adapt these algorithms, showing significant potential for improving training stability through effective noise management.


\subsection{Hyper-parameter Sensitivity}
\label{subsec: hypersens}

\begin{figure}
	\centering
	\subfigure[The weight of cluster-aware loss]{\label{subfig:lambda}
		\includegraphics[width=0.23\textwidth]{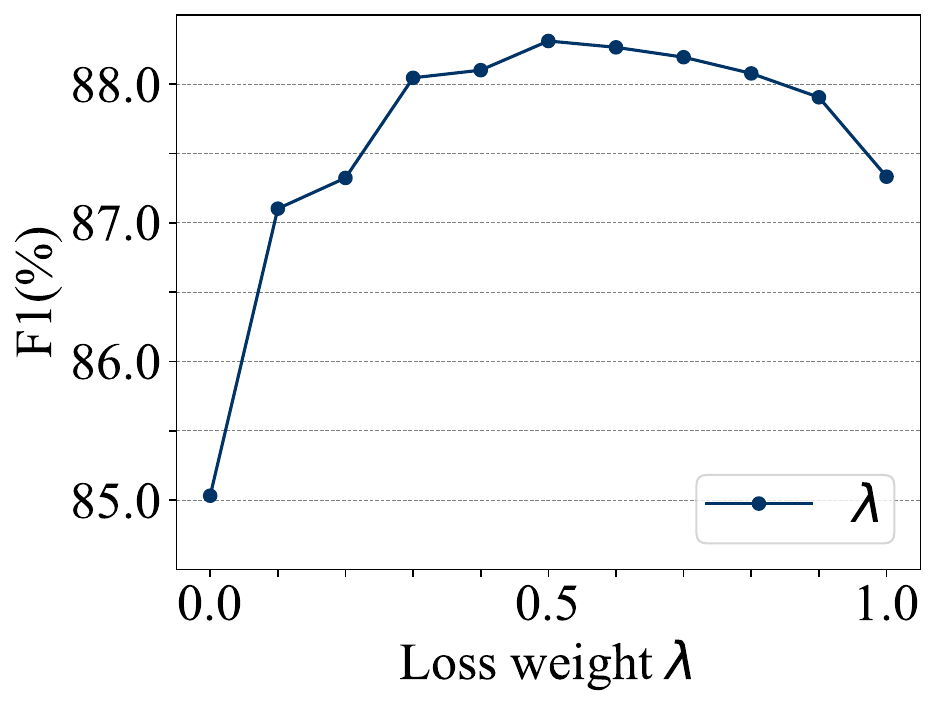}
	}
	\hspace{-0.1in}
	\subfigure[Parameters of DBSCAN]{\label{subfig:dbscan}
		\includegraphics[width=0.23\textwidth]{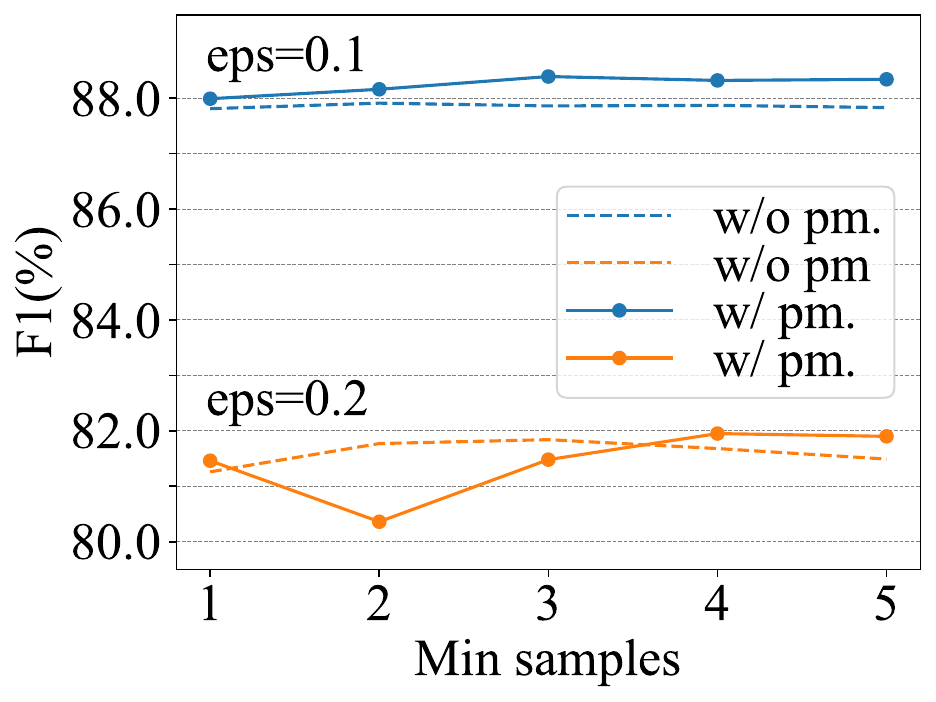}
	}
	\vspace{-3pt}
	\caption{\label{fig:hyperp} Analysis of hyper-parameters.}
	\vspace{-3pt}
\end{figure}

In this subsection, we investigate the performance variation when adjusting main hyper-parameters in \model.

\vpara{Sensitivity of the weight of cluster-aware loss.} We examined how the parameter \(\lambda\) impacts name disambiguation performance in the range of $[0, 1]$. The results in Figure \ref{subfig:lambda} indicate that the best \(\lambda\) value is 0.5, striking a balance between local linkage learning and cluster-aware learning.  A larger $\lambda$ approaching $1$ yields better results than $\lambda \to 0$,
demonstrating the effectiveness of our proposed end-to-end cluster-aware learning component.

\vpara{Parameters of DBSCAN.}
The maximum distance between neighboring samples $eps$, 
and the minimum samples in a neighborhood
$min\_samples$, can both impact the performance of DBSCAN, as illustrated in Figure \ref{subfig:dbscan}. 
Our observations reveal that $eps$ has a more pronounced impact on performance, and reducing it from 0.2 to 0.1 leads to a significant improvement,
implying that the strict restriction of neighboring distance would generate better clustering results. 
The relationship between $min\_samples$ and post-match is intertwined. 
As demonstrated by the line with circle dots, performance enhances as $min\_samples$ increases from 1 to 5, resulting in more outliers that could be addressed by post-match strategies.


\section{Conclusion}
\label{sec: con}

In this work, we introduce the first attempt to address the from-scratch name disambiguation problem by mutually enhancing the local and global optimal signals within an end-to-end framework. Specifically, our global clustering task utilizes local pairwise similarities to create pseudo-clustering outcomes, and these global optimization signals are used as feedback to further refine the local pairwise characteristics. Our extensive experiments validate the effectiveness of each component in our proposed framework. In the future, we aim to mitigate inherent biases in different author names and explore commonalities across various names by leveraging extensive disambiguation data and large language models.

\begin{acks}

This work was supported by Technology and Innovation Major Project of the Ministry of Science and Technology of China under Grant 2020AAA0108400, NSFC for Distinguished Young Scholar 61825602, and the New Cornerstone Science Foundation through the XPLORER PRIZE.

\end{acks}

\clearpage

\bibliographystyle{ACM-Reference-Format}
\balance
\bibliography{ref} 


\begin{thebibliography}{42}


\ifx \showCODEN    \undefined \def \showCODEN     #1{\unskip}     \fi
\ifx \showDOI      \undefined \def \showDOI       #1{#1}\fi
\ifx \showISBNx    \undefined \def \showISBNx     #1{\unskip}     \fi
\ifx \showISBNxiii \undefined \def \showISBNxiii  #1{\unskip}     \fi
\ifx \showISSN     \undefined \def \showISSN      #1{\unskip}     \fi
\ifx \showLCCN     \undefined \def \showLCCN      #1{\unskip}     \fi
\ifx \shownote     \undefined \def \shownote      #1{#1}          \fi
\ifx \showarticletitle \undefined \def \showarticletitle #1{#1}   \fi
\ifx \showURL      \undefined \def \showURL       {\relax}        \fi
\providecommand\bibfield[2]{#2}
\providecommand\bibinfo[2]{#2}
\providecommand\natexlab[1]{#1}
\providecommand\showeprint[2][]{arXiv:#2}

\bibitem[Ankerst et~al\mbox{.}(1999)]%
        {ankerst1999optics}
\bibfield{author}{\bibinfo{person}{Mihael Ankerst}, \bibinfo{person}{Markus~M Breunig}, \bibinfo{person}{Hans-Peter Kriegel}, {and} \bibinfo{person}{J{\"o}rg Sander}.} \bibinfo{year}{1999}\natexlab{}.
\newblock \showarticletitle{{OPTICS:} Ordering Points To Identify the Clustering Structure}. In \bibinfo{booktitle}{\emph{Proceedings {ACM} {SIGMOD} International Conference on Management of Data}}. \bibinfo{pages}{49--60}.
\newblock
\urldef\tempurl%
\url{https://doi.org/10.1145/304182.304187}
\showDOI{\tempurl}


\bibitem[Atarashi et~al\mbox{.}(2017)]%
        {atarashi2017deep}
\bibfield{author}{\bibinfo{person}{Kyohei Atarashi}, \bibinfo{person}{Satoshi Oyama}, \bibinfo{person}{Masahito Kurihara}, {and} \bibinfo{person}{Kazune Furudo}.} \bibinfo{year}{2017}\natexlab{}.
\newblock \showarticletitle{A deep neural network for pairwise classification: Enabling feature conjunctions and ensuring symmetry}. In \bibinfo{booktitle}{\emph{Advances in Knowledge Discovery and Data Mining: 21st Pacific-Asia Conference}}. \bibinfo{pages}{83--95}.
\newblock
\urldef\tempurl%
\url{https://doi.org/10.1007/978-3-319-57454-7_7}
\showDOI{\tempurl}


\bibitem[Beltagy et~al\mbox{.}(2019)]%
        {Beltagy2019SciBERT}
\bibfield{author}{\bibinfo{person}{Iz Beltagy}, \bibinfo{person}{Kyle Lo}, {and} \bibinfo{person}{Arman Cohan}.} \bibinfo{year}{2019}\natexlab{}.
\newblock \showarticletitle{SciBERT: {A} Pretrained Language Model for Scientific Text}. In \bibinfo{booktitle}{\emph{Proceedings of the 2019 Conference on Empirical Methods in Natural Language Processing and the 9th International Joint Conference on Natural Language Processing}}. \bibinfo{pages}{3613--3618}.
\newblock
\urldef\tempurl%
\url{https://doi.org/10.18653/V1/D19-1371}
\showDOI{\tempurl}


\bibitem[Cen et~al\mbox{.}(2013)]%
        {cen2013author}
\bibfield{author}{\bibinfo{person}{Lei Cen}, \bibinfo{person}{Eduard~C Dragut}, \bibinfo{person}{Luo Si}, {and} \bibinfo{person}{Mourad Ouzzani}.} \bibinfo{year}{2013}\natexlab{}.
\newblock \showarticletitle{Author disambiguation by hierarchical agglomerative clustering with adaptive stopping criterion}. In \bibinfo{booktitle}{\emph{Proceedings of the 36th International ACM SIGIR conference on Research and development in information retrieval}}. \bibinfo{pages}{741--744}.
\newblock
\urldef\tempurl%
\url{https://doi.org/10.1145/2484028.2484157}
\showDOI{\tempurl}


\bibitem[Chen et~al\mbox{.}(2023)]%
        {chen2023web}
\bibfield{author}{\bibinfo{person}{Bo Chen}, \bibinfo{person}{Jing Zhang}, \bibinfo{person}{Fanjin Zhang}, \bibinfo{person}{Tianyi Han}, \bibinfo{person}{Yuqing Cheng}, \bibinfo{person}{Xiaoyan Li}, \bibinfo{person}{Yuxiao Dong}, {and} \bibinfo{person}{Jie Tang}.} \bibinfo{year}{2023}\natexlab{}.
\newblock \showarticletitle{Web-Scale Academic Name Disambiguation: The WhoIsWho Benchmark, Leaderboard, and Toolkit}. In \bibinfo{booktitle}{\emph{Proceedings of the 29th ACM SIGKDD Conference on Knowledge Discovery and Data Mining}}. \bibinfo{pages}{3817–3828}.
\newblock
\urldef\tempurl%
\url{https://doi.org/10.1145/3580305.3599930}
\showDOI{\tempurl}


\bibitem[Chen et~al\mbox{.}(2021)]%
        {chen2021name}
\bibfield{author}{\bibinfo{person}{Ya Chen}, \bibinfo{person}{Hongliang Yuan}, \bibinfo{person}{Tingting Liu}, {and} \bibinfo{person}{Nan Ding}.} \bibinfo{year}{2021}\natexlab{}.
\newblock \showarticletitle{Name disambiguation based on graph convolutional network}.
\newblock \bibinfo{journal}{\emph{Scientific Programming}}  \bibinfo{volume}{2021} (\bibinfo{year}{2021}), \bibinfo{pages}{1--11}.
\newblock
\urldef\tempurl%
\url{https://doi.org/10.1155/2021/5577692}
\showDOI{\tempurl}


\bibitem[Devlin et~al\mbox{.}({[n.\,d.]})]%
        {devlin2018bert}
\bibfield{author}{\bibinfo{person}{Jacob Devlin}, \bibinfo{person}{Ming-Wei Chang}, \bibinfo{person}{Kenton Lee}, {and} \bibinfo{person}{Kristina Toutanova}.} \bibinfo{year}{[n.\,d.]}\natexlab{}.
\newblock \showarticletitle{{BERT}: Pre-training of Deep Bidirectional Transformers for Language Understanding}. In \bibinfo{booktitle}{\emph{Proceedings of the 2019 Conference of the North {A}merican Chapter of the Association for Computational Linguistics: Human Language Technologies, Volume 1 (Long and Short Papers)}}. \bibinfo{pages}{4171--4186}.
\newblock
\urldef\tempurl%
\url{https://doi.org/10.18653/v1/N19-1423}
\showDOI{\tempurl}


\bibitem[Ester et~al\mbox{.}(1996)]%
        {ester1996density}
\bibfield{author}{\bibinfo{person}{Martin Ester}, \bibinfo{person}{Hans-Peter Kriegel}, \bibinfo{person}{J{\"o}rg Sander}, \bibinfo{person}{Xiaowei Xu}, {et~al\mbox{.}}} \bibinfo{year}{1996}\natexlab{}.
\newblock \showarticletitle{A density-based algorithm for discovering clusters in large spatial databases with noise}. In \bibinfo{booktitle}{\emph{Proceedings of the Second International Conference on Knowledge Discovery and Data Mining}}. \bibinfo{pages}{226–231}.
\newblock


\bibitem[Fan et~al\mbox{.}(2011)]%
        {fan2011graph}
\bibfield{author}{\bibinfo{person}{Xiaoming Fan}, \bibinfo{person}{Jianyong Wang}, \bibinfo{person}{Xu Pu}, \bibinfo{person}{Lizhu Zhou}, {and} \bibinfo{person}{Bing Lv}.} \bibinfo{year}{2011}\natexlab{}.
\newblock \showarticletitle{On graph-based name disambiguation}.
\newblock \bibinfo{journal}{\emph{Journal of Data and Information Quality}} \bibinfo{volume}{2}, \bibinfo{number}{2} (\bibinfo{year}{2011}), \bibinfo{pages}{1--23}.
\newblock
\urldef\tempurl%
\url{https://doi.org/10.1145/1891879.1891883}
\showDOI{\tempurl}


\bibitem[Frey and Dueck(2007)]%
        {frey2007clustering}
\bibfield{author}{\bibinfo{person}{Brendan~J Frey} {and} \bibinfo{person}{Delbert Dueck}.} \bibinfo{year}{2007}\natexlab{}.
\newblock \showarticletitle{Clustering by passing messages between data points}.
\newblock \bibinfo{journal}{\emph{science}} \bibinfo{volume}{315}, \bibinfo{number}{5814} (\bibinfo{year}{2007}), \bibinfo{pages}{972--976}.
\newblock
\urldef\tempurl%
\url{https://doi.org/10.1126/science.1136800}
\showDOI{\tempurl}


\bibitem[Glorot and Bengio(2010)]%
        {glorot2010understanding}
\bibfield{author}{\bibinfo{person}{Xavier Glorot} {and} \bibinfo{person}{Yoshua Bengio}.} \bibinfo{year}{2010}\natexlab{}.
\newblock \showarticletitle{Understanding the difficulty of training deep feedforward neural networks}. In \bibinfo{booktitle}{\emph{Proceedings of the Thirteenth International Conference on Artificial Intelligence and Statistics}}, Vol.~\bibinfo{volume}{9}. \bibinfo{pages}{249--256}.
\newblock


\bibitem[Han et~al\mbox{.}(2005)]%
        {han2005name}
\bibfield{author}{\bibinfo{person}{Hui Han}, \bibinfo{person}{Hongyuan Zha}, {and} \bibinfo{person}{C~Lee Giles}.} \bibinfo{year}{2005}\natexlab{}.
\newblock \showarticletitle{Name disambiguation in author citations using a k-way spectral clustering method}. In \bibinfo{booktitle}{\emph{Proceedings of the 5th ACM/IEEE-CS Joint Conference on Digital Libraries}}. \bibinfo{pages}{334–343}.
\newblock
\urldef\tempurl%
\url{https://doi.org/10.1145/1065385.1065462}
\showDOI{\tempurl}


\bibitem[Hearst et~al\mbox{.}(1998)]%
        {hearst1998support}
\bibfield{author}{\bibinfo{person}{Marti~A. Hearst}, \bibinfo{person}{Susan~T Dumais}, \bibinfo{person}{Edgar Osuna}, \bibinfo{person}{John Platt}, {and} \bibinfo{person}{Bernhard Scholkopf}.} \bibinfo{year}{1998}\natexlab{}.
\newblock \showarticletitle{Support vector machines}.
\newblock \bibinfo{journal}{\emph{IEEE Intelligent Systems and their applications}} \bibinfo{volume}{13}, \bibinfo{number}{4} (\bibinfo{year}{1998}), \bibinfo{pages}{18--28}.
\newblock
\urldef\tempurl%
\url{https://doi.org/10.1109/5254.708428}
\showDOI{\tempurl}


\bibitem[Heller and Ghahramani(2005)]%
        {heller2005bayesian}
\bibfield{author}{\bibinfo{person}{Katherine~A Heller} {and} \bibinfo{person}{Zoubin Ghahramani}.} \bibinfo{year}{2005}\natexlab{}.
\newblock \showarticletitle{Bayesian hierarchical clustering}. In \bibinfo{booktitle}{\emph{Proceedings of the 22nd International Conference on Machine Learning}}. \bibinfo{pages}{297–304}.
\newblock
\urldef\tempurl%
\url{https://doi.org/10.1145/1102351.1102389}
\showDOI{\tempurl}


\bibitem[Kingma and Ba(2014)]%
        {kingma2014adam}
\bibfield{author}{\bibinfo{person}{Diederik~P Kingma} {and} \bibinfo{person}{Jimmy Ba}.} \bibinfo{year}{2014}\natexlab{}.
\newblock \showarticletitle{Adam: A method for stochastic optimization}.
\newblock \bibinfo{journal}{\emph{arXiv preprint arXiv:1412.6980}} (\bibinfo{year}{2014}).
\newblock
\urldef\tempurl%
\url{https://doi.org/10.48550/arXiv.1412.6980}
\showDOI{\tempurl}


\bibitem[Kipf and Welling(2016)]%
        {kipf2016variational}
\bibfield{author}{\bibinfo{person}{Thomas~N. Kipf} {and} \bibinfo{person}{Max Welling}.} \bibinfo{year}{2016}\natexlab{}.
\newblock \showarticletitle{Variational graph auto-encoders}.
\newblock \bibinfo{journal}{\emph{arXiv preprint arXiv:1611.07308}} (\bibinfo{year}{2016}).
\newblock
\urldef\tempurl%
\url{https://doi.org/10.48550/arXiv.1611.07308}
\showDOI{\tempurl}


\bibitem[Kipf and Welling(2017)]%
        {kipf2016semi}
\bibfield{author}{\bibinfo{person}{Thomas~N. Kipf} {and} \bibinfo{person}{Max Welling}.} \bibinfo{year}{2017}\natexlab{}.
\newblock \showarticletitle{Semi-Supervised Classification with Graph Convolutional Networks}.
\newblock  (\bibinfo{year}{2017}).
\newblock


\bibitem[Kristiadi et~al\mbox{.}(2019)]%
        {kristiadi2019incorporating}
\bibfield{author}{\bibinfo{person}{Agustinus Kristiadi}, \bibinfo{person}{Mohammad~Asif Khan}, \bibinfo{person}{Denis Lukovnikov}, \bibinfo{person}{Jens Lehmann}, {and} \bibinfo{person}{Asja Fischer}.} \bibinfo{year}{2019}\natexlab{}.
\newblock \showarticletitle{Incorporating literals into knowledge graph embeddings}. In \bibinfo{booktitle}{\emph{The Semantic Web -- ISWC 2019}}. \bibinfo{pages}{347--363}.
\newblock
\urldef\tempurl%
\url{https://doi.org/10.1007/978-3-030-30793-6_20}
\showDOI{\tempurl}


\bibitem[Li et~al\mbox{.}(2021)]%
        {li2021disambiguating}
\bibfield{author}{\bibinfo{person}{Na Li}, \bibinfo{person}{Renyu Zhu}, \bibinfo{person}{Xiaoxu Zhou}, \bibinfo{person}{Xiangnan He}, \bibinfo{person}{Wenyuan Cai}, \bibinfo{person}{Ming Gao}, {and} \bibinfo{person}{Aoying Zhou}.} \bibinfo{year}{2021}\natexlab{}.
\newblock \showarticletitle{On disambiguating authors: Collaboration network reconstruction in a bottom-up manner}. In \bibinfo{booktitle}{\emph{2021 IEEE 37th International Conference on Data Engineering}}. \bibinfo{pages}{888--899}.
\newblock


\bibitem[Liu et~al\mbox{.}(2022)]%
        {liu2022oag}
\bibfield{author}{\bibinfo{person}{Xiao Liu}, \bibinfo{person}{Da Yin}, \bibinfo{person}{Jingnan Zheng}, \bibinfo{person}{Xingjian Zhang}, \bibinfo{person}{Peng Zhang}, \bibinfo{person}{Hongxia Yang}, \bibinfo{person}{Yuxiao Dong}, {and} \bibinfo{person}{Jie Tang}.} \bibinfo{year}{2022}\natexlab{}.
\newblock \showarticletitle{OAG-BERT: Towards a Unified Backbone Language Model for Academic Knowledge Services}. In \bibinfo{booktitle}{\emph{Proceedings of the 28th ACM SIGKDD Conference on Knowledge Discovery and Data Mining}}. \bibinfo{pages}{3418–3428}.
\newblock
\urldef\tempurl%
\url{https://doi.org/10.1145/3534678.3539210}
\showDOI{\tempurl}


\bibitem[Louppe et~al\mbox{.}(2016)]%
        {louppe2016ethnicity}
\bibfield{author}{\bibinfo{person}{Gilles Louppe}, \bibinfo{person}{Hussein~T Al-Natsheh}, \bibinfo{person}{Mateusz Susik}, {and} \bibinfo{person}{Eamonn~James Maguire}.} \bibinfo{year}{2016}\natexlab{}.
\newblock \showarticletitle{Ethnicity sensitive author disambiguation using semi-supervised learning}. In \bibinfo{booktitle}{\emph{Knowledge Engineering and Semantic Web: 7th International Conference}}. \bibinfo{pages}{272--287}.
\newblock
\urldef\tempurl%
\url{https://doi.org/10.1007/978-3-319-45880-9_21}
\showDOI{\tempurl}


\bibitem[McInnes and Healy(2017)]%
        {mcinnes2017accelerated}
\bibfield{author}{\bibinfo{person}{Leland McInnes} {and} \bibinfo{person}{John Healy}.} \bibinfo{year}{2017}\natexlab{}.
\newblock \showarticletitle{Accelerated hierarchical density based clustering}. In \bibinfo{booktitle}{\emph{2017 IEEE International Conference on Data Mining Workshops}}. \bibinfo{pages}{33--42}.
\newblock
\urldef\tempurl%
\url{https://doi.org/10.1109/ICDMW.2017.12}
\showDOI{\tempurl}


\bibitem[Mikolov et~al\mbox{.}(2013)]%
        {mikolov2013efficient}
\bibfield{author}{\bibinfo{person}{Tom{\'{a}}s Mikolov}, \bibinfo{person}{Kai Chen}, \bibinfo{person}{Greg Corrado}, {and} \bibinfo{person}{Jeffrey Dean}.} \bibinfo{year}{2013}\natexlab{}.
\newblock \showarticletitle{Efficient Estimation of Word Representations in Vector Space}. In \bibinfo{booktitle}{\emph{1st International Conference on Learning Representations}}.
\newblock
\urldef\tempurl%
\url{http://arxiv.org/abs/1301.3781}
\showURL{%
\tempurl}


\bibitem[M{\"u}llner(2011)]%
        {mullner2011modern}
\bibfield{author}{\bibinfo{person}{Daniel M{\"u}llner}.} \bibinfo{year}{2011}\natexlab{}.
\newblock \showarticletitle{Modern hierarchical, agglomerative clustering algorithms}.
\newblock \bibinfo{journal}{\emph{arXiv preprint arXiv:1109.2378}} (\bibinfo{year}{2011}).
\newblock
\urldef\tempurl%
\url{https://doi.org/10.48550/arXiv.1109.2378}
\showDOI{\tempurl}


\bibitem[On et~al\mbox{.}(2012)]%
        {on2012scalable}
\bibfield{author}{\bibinfo{person}{Byung-Won On}, \bibinfo{person}{Ingyu Lee}, {and} \bibinfo{person}{Dongwon Lee}.} \bibinfo{year}{2012}\natexlab{}.
\newblock \showarticletitle{Scalable clustering methods for the name disambiguation problem}.
\newblock \bibinfo{journal}{\emph{Knowledge and Information Systems}}  \bibinfo{volume}{31} (\bibinfo{year}{2012}), \bibinfo{pages}{129--151}.
\newblock
\urldef\tempurl%
\url{https://doi.org/10.1007/s10115-011-0397-1}
\showDOI{\tempurl}


\bibitem[Pooja et~al\mbox{.}(2021)]%
        {pooja2021exploiting}
\bibfield{author}{\bibinfo{person}{KM Pooja}, \bibinfo{person}{Samrat Mondal}, {and} \bibinfo{person}{Joydeep Chandra}.} \bibinfo{year}{2021}\natexlab{}.
\newblock \showarticletitle{Exploiting similarities across multiple dimensions for author name disambiguation}.
\newblock \bibinfo{journal}{\emph{Scientometrics}}  \bibinfo{volume}{126} (\bibinfo{year}{2021}), \bibinfo{pages}{7525--7560}.
\newblock
\urldef\tempurl%
\url{https://doi.org/10.1007/s11192-021-04101-y}
\showDOI{\tempurl}


\bibitem[Pooja et~al\mbox{.}(2022)]%
        {pooja2022exploiting}
\bibfield{author}{\bibinfo{person}{Km Pooja}, \bibinfo{person}{Samrat Mondal}, {and} \bibinfo{person}{Joydeep Chandra}.} \bibinfo{year}{2022}\natexlab{}.
\newblock \showarticletitle{Exploiting Higher Order Multi-dimensional Relationships with Self-attention for Author Name Disambiguation}.
\newblock \bibinfo{journal}{\emph{ACM Transactions on Knowledge Discovery from Data}} \bibinfo{volume}{16}, \bibinfo{number}{5} (\bibinfo{year}{2022}), \bibinfo{pages}{1--23}.
\newblock
\urldef\tempurl%
\url{https://doi.org/10.1145/3502730}
\showDOI{\tempurl}


\bibitem[Qiao et~al\mbox{.}(2019)]%
        {qiao2019unsupervised}
\bibfield{author}{\bibinfo{person}{Ziyue Qiao}, \bibinfo{person}{Yi Du}, \bibinfo{person}{Yanjie Fu}, \bibinfo{person}{Pengfei Wang}, {and} \bibinfo{person}{Yuanchun Zhou}.} \bibinfo{year}{2019}\natexlab{}.
\newblock \showarticletitle{Unsupervised author disambiguation using heterogeneous graph convolutional network embedding}. In \bibinfo{booktitle}{\emph{2019 IEEE international conference on big data}}. \bibinfo{pages}{910--919}.
\newblock
\urldef\tempurl%
\url{https://doi.org/10.1109/BigData47090.2019.9005458}
\showDOI{\tempurl}


\bibitem[Santini et~al\mbox{.}(2022)]%
        {santini2022knowledge}
\bibfield{author}{\bibinfo{person}{Cristian Santini}, \bibinfo{person}{Genet~Asefa Gesese}, \bibinfo{person}{Silvio Peroni}, \bibinfo{person}{Aldo Gangemi}, \bibinfo{person}{Harald Sack}, {and} \bibinfo{person}{Mehwish Alam}.} \bibinfo{year}{2022}\natexlab{}.
\newblock \showarticletitle{A knowledge graph embeddings based approach for author name disambiguation using literals}.
\newblock \bibinfo{journal}{\emph{Scientometrics}} \bibinfo{volume}{127}, \bibinfo{number}{8} (\bibinfo{year}{2022}), \bibinfo{pages}{4887--4912}.
\newblock
\urldef\tempurl%
\url{https://doi.org/10.1007/s11192-022-04426-2}
\showDOI{\tempurl}


\bibitem[Shin et~al\mbox{.}(2014)]%
        {shin2014author}
\bibfield{author}{\bibinfo{person}{Dongwook Shin}, \bibinfo{person}{Taehwan Kim}, \bibinfo{person}{Joongmin Choi}, {and} \bibinfo{person}{Jungsun Kim}.} \bibinfo{year}{2014}\natexlab{}.
\newblock \showarticletitle{Author name disambiguation using a graph model with node splitting and merging based on bibliographic information}.
\newblock \bibinfo{journal}{\emph{Scientometrics}}  \bibinfo{volume}{100} (\bibinfo{year}{2014}), \bibinfo{pages}{15--50}.
\newblock
\urldef\tempurl%
\url{https://doi.org/10.1007/s11192-014-1289-4}
\showDOI{\tempurl}


\bibitem[Sun et~al\mbox{.}(2020)]%
        {sun2020pairwise}
\bibfield{author}{\bibinfo{person}{Qingyun Sun}, \bibinfo{person}{Hao Peng}, \bibinfo{person}{Jianxin Li}, \bibinfo{person}{Senzhang Wang}, \bibinfo{person}{Xiangyun Dong}, \bibinfo{person}{Liangxuan Zhao}, \bibinfo{person}{S~Yu Philip}, {and} \bibinfo{person}{Lifang He}.} \bibinfo{year}{2020}\natexlab{}.
\newblock \showarticletitle{Pairwise learning for name disambiguation in large-scale heterogeneous academic networks}. In \bibinfo{booktitle}{\emph{2020 IEEE International Conference on Data Mining}}. \bibinfo{pages}{511--520}.
\newblock
\urldef\tempurl%
\url{https://doi.org/10.1109/ICDM50108.2020.00060}
\showDOI{\tempurl}


\bibitem[Tang et~al\mbox{.}(2011)]%
        {tang2011unified}
\bibfield{author}{\bibinfo{person}{Jie Tang}, \bibinfo{person}{Alvis~CM Fong}, \bibinfo{person}{Bo Wang}, {and} \bibinfo{person}{Jing Zhang}.} \bibinfo{year}{2011}\natexlab{}.
\newblock \showarticletitle{A unified probabilistic framework for name disambiguation in digital library}.
\newblock \bibinfo{journal}{\emph{IEEE Transactions on Knowledge and Data Engineering}} \bibinfo{volume}{24}, \bibinfo{number}{6} (\bibinfo{year}{2011}), \bibinfo{pages}{975--987}.
\newblock
\urldef\tempurl%
\url{https://doi.org/10.1109/TKDE.2011.13.}
\showDOI{\tempurl}


\bibitem[Tang et~al\mbox{.}(2008)]%
        {tang2008arnetminer}
\bibfield{author}{\bibinfo{person}{Jie Tang}, \bibinfo{person}{Jing Zhang}, \bibinfo{person}{Limin Yao}, \bibinfo{person}{Juanzi Li}, \bibinfo{person}{Li Zhang}, {and} \bibinfo{person}{Zhong Su}.} \bibinfo{year}{2008}\natexlab{}.
\newblock \showarticletitle{Arnetminer: extraction and mining of academic social networks}. In \bibinfo{booktitle}{\emph{Proceedings of the 14th ACM SIGKDD international conference on Knowledge discovery and data mining}}. \bibinfo{pages}{990–998}.
\newblock
\urldef\tempurl%
\url{https://doi.org/10.1145/1401890.1402008}
\showDOI{\tempurl}


\bibitem[Tang and Walsh(2010)]%
        {tang2010bibliometric}
\bibfield{author}{\bibinfo{person}{Li Tang} {and} \bibinfo{person}{John Walsh}.} \bibinfo{year}{2010}\natexlab{}.
\newblock \showarticletitle{Bibliometric fingerprints: name disambiguation based on approximate structure equivalence of cognitive maps}.
\newblock \bibinfo{journal}{\emph{Scientometrics}} \bibinfo{volume}{84}, \bibinfo{number}{3} (\bibinfo{year}{2010}), \bibinfo{pages}{763--784}.
\newblock
\urldef\tempurl%
\url{https://doi.org/10.1007/s11192-010-0196-6}
\showDOI{\tempurl}


\bibitem[Veli{\v{c}}kovi{\'{c}} et~al\mbox{.}(2018)]%
        {velivckovic2018graph}
\bibfield{author}{\bibinfo{person}{Petar Veli{\v{c}}kovi{\'{c}}}, \bibinfo{person}{Guillem Cucurull}, \bibinfo{person}{Arantxa Casanova}, \bibinfo{person}{Adriana Romero}, \bibinfo{person}{Pietro Li{\`{o}}}, {and} \bibinfo{person}{Yoshua Bengio}.} \bibinfo{year}{2018}\natexlab{}.
\newblock \showarticletitle{{Graph Attention Networks}}.
\newblock \bibinfo{journal}{\emph{International Conference on Learning Representations}} (\bibinfo{year}{2018}).
\newblock
\urldef\tempurl%
\url{https://openreview.net/forum?id=rJXMpikCZ}
\showURL{%
\tempurl}


\bibitem[Xie et~al\mbox{.}(2016)]%
        {xie2016unsupervised}
\bibfield{author}{\bibinfo{person}{Junyuan Xie}, \bibinfo{person}{Ross~B. Girshick}, {and} \bibinfo{person}{Ali Farhadi}.} \bibinfo{year}{2016}\natexlab{}.
\newblock \showarticletitle{Unsupervised Deep Embedding for Clustering Analysis}. In \bibinfo{booktitle}{\emph{Proceedings of the 33nd International Conference on Machine Learning}}, Vol.~\bibinfo{volume}{48}. \bibinfo{pages}{478--487}.
\newblock
\urldef\tempurl%
\url{http://proceedings.mlr.press/v48/xieb16.html}
\showURL{%
\tempurl}


\bibitem[Xu et~al\mbox{.}(2019)]%
        {xu2018powerful}
\bibfield{author}{\bibinfo{person}{Keyulu Xu}, \bibinfo{person}{Weihua Hu}, \bibinfo{person}{Jure Leskovec}, {and} \bibinfo{person}{Stefanie Jegelka}.} \bibinfo{year}{2019}\natexlab{}.
\newblock \showarticletitle{How Powerful are Graph Neural Networks?}. In \bibinfo{booktitle}{\emph{International Conference on Learning Representations}}.
\newblock
\urldef\tempurl%
\url{https://openreview.net/forum?id=ryGs6iA5Km}
\showURL{%
\tempurl}


\bibitem[Yoshida et~al\mbox{.}(2010)]%
        {yoshida2010person}
\bibfield{author}{\bibinfo{person}{Minoru Yoshida}, \bibinfo{person}{Masaki Ikeda}, \bibinfo{person}{Shingo Ono}, \bibinfo{person}{Issei Sato}, {and} \bibinfo{person}{Hiroshi Nakagawa}.} \bibinfo{year}{2010}\natexlab{}.
\newblock \showarticletitle{Person name disambiguation by bootstrapping}. In \bibinfo{booktitle}{\emph{Proceedings of the 33rd International ACM SIGIR Conference on Research and Development in Information Retrieval}}. \bibinfo{pages}{10–17}.
\newblock
\urldef\tempurl%
\url{https://doi.org/10.1145/1835449.1835454}
\showDOI{\tempurl}


\bibitem[Zhang and Al~Hasan(2017)]%
        {zhang2017name}
\bibfield{author}{\bibinfo{person}{Baichuan Zhang} {and} \bibinfo{person}{Mohammad Al~Hasan}.} \bibinfo{year}{2017}\natexlab{}.
\newblock \showarticletitle{Name disambiguation in anonymized graphs using network embedding}. In \bibinfo{booktitle}{\emph{Proceedings of the 2017 ACM on Conference on Information and Knowledge Management}}. \bibinfo{pages}{1239--1248}.
\newblock
\urldef\tempurl%
\url{https://doi.org/10.1145/3132847.3132873}
\showDOI{\tempurl}


\bibitem[Zhang et~al\mbox{.}(2019a)]%
        {zhang2019oag}
\bibfield{author}{\bibinfo{person}{Fanjin Zhang}, \bibinfo{person}{Xiao Liu}, \bibinfo{person}{Jie Tang}, \bibinfo{person}{Yuxiao Dong}, \bibinfo{person}{Peiran Yao}, \bibinfo{person}{Jie Zhang}, \bibinfo{person}{Xiaotao Gu}, \bibinfo{person}{Yan Wang}, \bibinfo{person}{Bin Shao}, \bibinfo{person}{Rui Li}, {et~al\mbox{.}}} \bibinfo{year}{2019}\natexlab{a}.
\newblock \showarticletitle{{OAG}: Toward linking large-scale heterogeneous entity graphs}. In \bibinfo{booktitle}{\emph{Proceedings of the 25th ACM SIGKDD International Conference on Knowledge Discovery \& Data Mining}}. \bibinfo{pages}{2585–2595}.
\newblock
\urldef\tempurl%
\url{https://doi.org/10.1145/3292500.3330785}
\showDOI{\tempurl}


\bibitem[Zhang et~al\mbox{.}(2019b)]%
        {zhang2019author}
\bibfield{author}{\bibinfo{person}{Wenjing Zhang}, \bibinfo{person}{Zhongmin Yan}, {and} \bibinfo{person}{Yongqing Zheng}.} \bibinfo{year}{2019}\natexlab{b}.
\newblock \showarticletitle{Author name disambiguation using graph node embedding method}. In \bibinfo{booktitle}{\emph{2019 IEEE 23rd international conference on computer supported cooperative work in design}}. \bibinfo{pages}{410--415}.
\newblock
\urldef\tempurl%
\url{https://doi.org/10.1109/CSCWD.2019.8791898}
\showDOI{\tempurl}


\bibitem[Zhang et~al\mbox{.}(2018)]%
        {zhang2018name}
\bibfield{author}{\bibinfo{person}{Yutao Zhang}, \bibinfo{person}{Fanjin Zhang}, \bibinfo{person}{Peiran Yao}, {and} \bibinfo{person}{Jie Tang}.} \bibinfo{year}{2018}\natexlab{}.
\newblock \showarticletitle{Name Disambiguation in AMiner: Clustering, Maintenance, and Human in the Loop}. In \bibinfo{booktitle}{\emph{Proceedings of the 24th ACM SIGKDD International Conference on Knowledge Discovery \& Data Mining}}. \bibinfo{pages}{1002–1011}.
\newblock
\urldef\tempurl%
\url{https://doi.org/10.1145/3219819.3219859}
\showDOI{\tempurl}


\end{thebibliography}


\appendix

\section{Appendices}
\subsection{Implementation Details of BOND}

In practice, our encoder is structured with two GAT layers, and the decoder employs inner product methodology for the graph auto-encoder. 
To identify optimal hidden layer dimensionalities, we explore a range of values from ${32, 64, 128, 256, 512}$. 
Similarly, the dimensionality of the fully-connected layer is examined within the set ${32, 64, 100, 256}$.
For the joint objective learning, the weight parameter \(\lambda\) for cluster-aware learning is fixed at $0.5$. 
All model parameters are initialized using the Xavier uniform distribution~\cite{glorot2010understanding} and optimized through the Adam optimizer~\cite{kingma2014adam}. 
Hyperparameters such as the learning rate and weight decay are systematically explored within the range of $1e^{-4}$ to $3e^{-3}$.
Each model associated with an author's name is meticulously trained over a course of $50$ epochs. 
All experiments are conducted on an NVIDIA GTX 3090Ti GPU.

\subsection{Graph Construction Methodology}

\vpara{Constructing Relational Edges}
In the preprocessing phase for establishing co-author relationships, our methodology is characterized by the normalization of author names (transforming name formats, for example, from "Li Jianrong" to "jianrongli") and the creation of connections between papers predicated on the intersection of their author lists, while explicitly excluding any authors subject to disambiguation. For the identification of co-venue relationships, our approach involves the conversion of venue names to lowercase, the elimination of stopwords, and the computation of overlaps to forge relational links.

In addressing co-organization relationships, which inherently display a higher susceptibility to noise, we employ the Jaccard Index. This measure is formally articulated as \( S = \frac {|p_a \cap p_b|}{|p_a| + |p_b| - |p_a \cap p_b|} \), wherein \( p_a \) and \( p_b \) represent the sets of words pertaining to the organizations from two disparate papers.

The experimental validation of these methodologies incorporates a range of combinations, from which we iteratively select the combination that yields the highest efficacy score for each type of relationship. As demonstrated in Table \ref{tb: edge}, preliminary analyses have revealed that the implementation of word overlap significantly enhances the identification capabilities for both co-author and co-venue relationships. In contrast, the Jaccard Index has been demonstrated to be particularly effective in attenuating the noise that is commonly associated with co-organization relationships. These findings underscore the nuanced efficacy of our preprocessing strategies in facilitating the accurate delineation of academic relationships.

\vpara{Threshold Strategies}
In our examination, we concentrated on the implications of adjusting threshold CoA, CoV, and CoO relationships. The calculation of CoA and CoV is based on word overlap, possessing a minimum threshold of 0. Conversely, CoO is evaluated utilizing the Jaccard Index, which exhibits a range from 0 to 1. As shown in Table \ref{tb: threshold}, our systematic experimentation revealed that the optimal thresholds for CoA, CoO, and CoV stand at 0, 0.6, and 2, respectively. Notably, CoA emerged as the most indicative of author name information and exerted the most significant influence on the performance of the model, thereby establishing it as a pivotal parameter within our analysis.

Conversely, CoO and CoV demonstrated a lower sensitivity to variations in threshold levels, with their performance exhibiting minor fluctuations across diverse configurations. This nuanced comprehension of the influence exerted by threshold adjustments is paramount to the refinement and optimization of our model's efficacy.

\begin{table}[htb]
	\newcolumntype{?}{!{\vrule width 1pt}}
	\newcolumntype{C}{>{\centering\arraybackslash}p{4em}}
	\caption{
		\label{tb: edge}\textbf{Graph construction (\%)} \textmd{CoA: co-authorship, CoO: co-organization, CoV: co-venue, CoA+O: CoA and CoO edges, CoA+O+V: CoA ,CoO and CoV edges.}
	}
	\centering 
	\renewcommand\arraystretch{1.0}
	\begin{tabular}{@{~}l?@{~}*{1}{CCC}@{~}}
		\toprule
		\textbf{Methods} & \textbf{CoA} & \textbf{CoA+O} & \textbf{CoA+O+V} \\
		\midrule
            Word overlap  & \textbf{84.18}	& 86.39	& \textbf{88.34}\\
            Jaccard index & 76.71	& \textbf{88.32} & 86.83\\
		\bottomrule
	\end{tabular}
\end{table}

\begin{table}[htb]
	\newcolumntype{?}{!{\vrule width 1pt}}
	\newcolumntype{C}{>{\centering\arraybackslash}p{4em}}
	\caption{
		\label{tb: threshold}\textbf{Thresholds of multi-view edge combinations.} (\%). \textmd{Ts: Thresholds. CoA and CoV, calculated by word overlap, have a minimum value of 0; CoO, measured using the Jaccard Index, ranges between 0 and 1. The CoA threshold of "0" means we will build an edge when CoA > 0. CoA+O threshold of "0, 0.2" implies that an edge is formed only when CoA > 0 and CoO > 0.2. }
	}
	\centering 
	\renewcommand\arraystretch{1.0}
	\begin{tabular}{@{~}l?@{~}*{1}{CCCCC}@{~}}
		\toprule
		\textbf{Ts} & \textbf{CoA} & \textbf{Ts} & \textbf{CoA+O} & \textbf{Ts} & \textbf{CoA+O+V} \\
		\midrule
            0&	\textbf{84.17}&	0, 0.2&	74.65&	0, 0.6, 0&	85.61\\
            1&	76.90&	0, 0.3&	84.95&	0, 0.6, 1&	87.83\\
            2&	71.59&	0, 0.4&	87.88&	0, 0.6, 2&	\textbf{88.34}\\
            3&	69.10&	0, 0.5&	87.88&	0, 0.6, 3&	88.13\\
            4&	65.62&	0, 0.6&	\textbf{88.30}&	0, 0.6, 4&	88.15\\
            5&	70.83&	0, 0.7&	88.02&	0, 0.6, 5&	88.13\\
		\bottomrule
	\end{tabular}
\end{table}

\subsection{Analysis of GNN Encoders}

We employ GAT as the GNN encoder in our model. 
We also compare GAT with other popular GNN models, including GCN~\cite{kipf2016semi} and GIN ~\cite{xu2018powerful}.
As illustrated in Table \ref{tb: gnn}, GAT exhibits superior performance compared to GCN and GIN, achieving 1.03\% improvement over GCN and 8.06\% improvement over GIN w.r.t. pairwise F1.
This is attributed to GAT's ability to assign adaptive importance to different edges through its attention mechanism. 
This feature mitigates the limitations associated with the unified type of edge, while simultaneously maintaining high efficiency.

\begin{table}[htb]
	\newcolumntype{?}{!{\vrule width 1pt}}
	\newcolumntype{C}{>{\centering\arraybackslash}p{4em}}
	\caption{
		\label{tb: gnn}GNN encoder (\%).
	}
	\centering 
	\renewcommand\arraystretch{1.0}
	\begin{tabular}{@{~}l?@{~}*{1}{CCC}@{~}}
		\toprule
		\textbf{Models} & \textbf{Precision} & \textbf{Recall} & \textbf{F1} \\
		\midrule
            GCN & 81.71 &	94.04 &     87.44\\
            GIN & 71.16	&   \textbf{96.0} &     81.75 \\
            GAT & \textbf{82.36}	&   95.25 & \textbf{88.34}\\
		\bottomrule
	\end{tabular}
\end{table}

\subsection{Out-Layer Size of the Fully Connected Layer}

In the training of the cluster-aware learning module, we utilize the transductive setting and dynamically adapt the size of the output layer within the fully connected layer based on the compression ratio multiplied by the number of nodes in the graph. As presented in Table \ref{tb: compress}, there is an observable performance enhancement of +3.55\% when the compression ratio is extended from 0.03 to 1.0. This outcome highlights the module's effectiveness in capturing the unique characteristics of individual name-associated graphs while accommodating the adaptability of the fully connected layers.

\begin{table}[htb]
	\newcolumntype{?}{!{\vrule width 1pt}}
	\newcolumntype{C}{>{\centering\arraybackslash}p{4em}}
	\caption{
		\label{tb: compress}Compression ratio (\%).
	}
	\centering 
	\renewcommand\arraystretch{1.0}
	\begin{tabular}{@{~}l?@{~}*{1}{CCC}@{~}}
		\toprule
		\textbf{Ratio} & \textbf{Precision} & \textbf{Recall} & \textbf{F1} \\
		\midrule
            0.03 & 75.67 & \textbf{95.31}  &	84.36   \\
            0.1  & 78.90 & 94.87  & 86.15  \\
            0.3  & 79.30 & 95.06  & 86.47  \\
            1.0  & \textbf{82.38} & 94.24  &	\textbf{87.91}  \\
            3.0  & 82.12 & 92.70  & 87.09   \\
		\bottomrule
	\end{tabular}
\end{table}

\subsection{Component-Wise Ablation Study}

We undertake an experimental investigation to elucidate the contributions of individual components within our model. The Feedforward Neural Network (FNN) serves as a comprehensive feature extractor, its training facilitated by pseudo labels generated through the DBSCAN clustering algorithm. The application of the inner product to the output of the FNN is instrumental in elucidating pairwise relationships among data points. 

Notably, the inner product emerges as a critical determinant of performance, evidencing its paramount importance in the precise capture of pairwise data relationships. Additionally, the FNN exhibits remarkable adaptability in adjusting embeddings to accommodate variations in graph sizes, a feature that is vital for the accurate disambiguation of names.

\begin{table}[htb]
	\newcolumntype{?}{!{\vrule width 1pt}}
	\newcolumntype{C}{>{\centering\arraybackslash}p{4em}}
	\caption{
		\label{tb: components}\textbf{Different components (\%).}
	}
	\centering 
	\renewcommand\arraystretch{1.0}
	\begin{tabular}{@{~}l?@{~}*{1}{CCC}@{~}}
		\toprule
		\textbf{Description} & \textbf{Precision} & \textbf{Recall} & \textbf{F1} \\
		\midrule
            W/o Inner product&	75.53&	\textbf{96.21}&	84.62 \\
            W/o DBSCAN&	77.58&	94.19&	85.08 \\
            W/o FNN&	\textbf{83.30}&	90.12&	86.58 \\
            \midrule
            \model(full)&	82.36&	95.25&	\textbf{88.34}\\
		\bottomrule
	\end{tabular}
\end{table}

\end{document}